\documentclass[prl,aps,twocolumn,superscriptaddress]{revtex4-2}

\usepackage{bm,bbm,color,float,dcolumn,amsmath,amssymb,graphicx,subfigure, mathtools, mhchem}

\usepackage[colorlinks=true]{hyperref}
\hypersetup{linkcolor=blue,citecolor=blue,urlcolor=blue}

\definecolor{psychedelicpurple}{rgb}{0.87, 0.0, 1.0}
\definecolor{violet}{rgb}{0.56, 0, 1}

\begin{document}

\title{Truncated Gaussian basis approach for simulating many-body dynamics}

\author{Nico Albert}
\affiliation{Institut f\"ur Theoretische Physik, Technische Universit\"at Dresden, 01062 Dresden, Germany}

\author{Yueshui Zhang}
\affiliation{Faculty of Physics and Arnold Sommerfeld Center for Theoretical Physics, Ludwig-Maximilians-Universit\"at M\"unchen, 80333 Munich, Germany}

\author{Hong-Hao Tu}
\email{h.tu@lmu.de}
\affiliation{Faculty of Physics and Arnold Sommerfeld Center for Theoretical Physics, Ludwig-Maximilians-Universit\"at M\"unchen, 80333 Munich, Germany}
\affiliation{Institut f\"ur Theoretische Physik, Technische Universit\"at Dresden, 01062 Dresden, Germany}

\date{\today}

\begin{abstract}
We propose a Truncated Gaussian Basis Approach (TGBA) for simulating the dynamics of quantum many-body systems. The approach constructs an effective Hamiltonian within a reduced subspace, spanned by fermionic Gaussian states, and diagonalizes it to obtain approximate eigenstates and eigenenergies. Symmetries can be exploited to perform parallel computation, enabling to simulate systems with much larger sizes. As an example, we compute the dynamic structure factor and study quench dynamics in a non-integrable quantum Ising chain, known as ``$E_8$ magnet''. The mass ratios calculated through the dynamic structure factor show excellent agreement with Zamolodchikov's analytical predictions. For quench dynamics we observe that time-evolving wave functions in the truncated subspace facilitates the simulation of long-time dynamics.
\end{abstract}

\maketitle

{\em Introduction} --- Efficient simulation of the dynamics of quantum many-body systems is an important yet challenging task in both experimental and theoretical physics~\cite{Plokovnikov2011,Georgescu2014,DAlessio2016,Abanin2019}. It serves not only as a key method for exploring novel quantum phases and phase transitions in many-body physics, but also as a vital tool for investigating fundamental problems that bridge quantum and statistical physics, such as quantum chaos and thermalization~\cite{DAlessio2016}. However, computational methods are often hindered by the exponential growth of the Hilbert space dimension as the system size increases or the rapid entanglement growth as time evolves, leading to significant difficulties in simulating dynamics in quantum many-body systems~\cite{Vidal2003,White2003,Daley2004,Haegeman2011,Haegeman2016,LiJW2024}.

The importance of simulating dynamics lies both in equilibrium and out of equilibrium scenarios. In equilibrium, real-frequency spectral functions contain crucial information about low-energy excitations and are directly linked to spectroscopic measurements. Due to the ill-conditioned nature of the analytic continuation~\cite{Jarrell1996,Sandvik1998,Fei2021} from the Matsubara frequency Green's functions, it is desirable to compute spectral functions directly in real frequency~\cite{White2008,Holzner2011,Nocera2016}. For non-equilibrium dynamics, significant advancements have been made in studying quench problems in cold atom experiments~\cite{Cheneau2012,Meinert2013,Langen2013,Kaufman2016}. However, as the initial state is typically far from the ground state, the entanglement entropy of time-evolved states often grows very rapidly with time (e.g., linear in time for one-dimensional critical systems~\cite{Calabrese2005}). This creates a substantial entanglement barrier, making it challenging to simulate long-time evolution in many-body systems with large sizes~\cite{Banuls2009, Mueller-Hermes2012, Frias-Perez2022, Lamb2024}.

In this work, we introduce a new algorithm, the Truncated Gaussian Basis Approach (TGBA), for simulating the dynamics in quantum many-body systems. The algorithm reduces the full Hilbert space by keeping a subspace spanned by fermionic Gaussian states~\footnote{The pure fermionic Gaussian states are also known as Hartree-Fock-Bogoliubov states. In the absence of pairing, they reduce to Hartree-Fock states.}. Within this subspace, fermionic Gaussian techniques enable us to diagonalize the truncated Hamiltonian and compute dynamical quantities, both in and out of equilibrium. Additionally, symmetries of the Hamiltonian, if available, can be fully exploited to perform parallel computations in different symmetry sectors. To demonstrate the power of our method, we apply it to the quantum Ising chain with both transverse and longitudinal fields. The dynamic structure factor is calculated and found to exhibit universal mass ratios which were predicted to emerge in the integrable $E_8$ Toda field theory~\cite{Zamolodchikov1989} and have also been observed experimentally in certain quasi-one-dimensional materials~\cite{Coldea2010,Amelin2020,ZhangZ2020,ZouHY2021,Amelin2022}. Furthermore, we demonstrate the capability of the TGBA to simulate non-equilibrium dynamics following a quantum quench. By employing an appropriate truncation scheme, we show that observables can be reliably computed up to long times. These results indicate that the TGBA is a promising method in simulating many-body dynamics.

{\em Method} --- In the TGBA, our main targets are Hamiltonians of the form $H = H_0 + V$, where $H_0$, the ``free part'', is (or can be mapped to) a fermionic quadratic Hamiltonian~\footnote{While $H_0$ is often part of $H$, one may also take $H_0$ as a ``mean-field'' approximation of $H$ and $V = H - H_0$.} so that one has full information about its spectrum $\{E_\alpha^{(0)}\}$ and eigenstates $|\phi_\alpha\rangle \in \mathcal{H}$ ($\mathcal{H}$: full Hilbert space). The eigenstates $|\phi_\alpha\rangle$ are (pure) fermionic Gaussian states~\cite{Bravyi2005}. Restricting to a suitable subspace $\mathcal{H}_\mathrm{trunc} \subset \mathcal{H}$, a variational ansatz for the eigenstates of the full Hamiltonian $H$ can be constructed from the eigenstates of $H_0$:
\begin{align}\label{eq:TGBA_ansatz}
    |\psi_\alpha \rangle = \sum_{|\phi_\beta\rangle \in \mathcal{H}_\mathrm{trunc}} M_{\alpha \beta} |\phi_\beta \rangle.
\end{align}
The optimal superposition coefficients $\{M_{\alpha \beta} \}$ and the approximated eigeneneriges $\{E_\alpha\}$ of the full Hamiltonian are determined by diagonalizing the effective Hamiltonian in the truncated subspace:
\begin{align}
    H_{\alpha \beta} \equiv \langle \phi_\alpha|H|\phi_\beta \rangle = E_\alpha^{(0)} \delta_{\alpha \beta} + \langle \phi_\alpha|V|\phi_\beta \rangle
\label{eq:trunc-ham}
\end{align}
with $|\phi_\alpha\rangle \in \mathcal{H}_{\mathrm{trunc}}$. If no truncation were performed, the method would be exact. However, in practice, one can only keep a maximum of $\chi$ states in $\mathcal{H}_\mathrm{trunc}$ and work with a $\chi$-dimensional \textit{truncated} Hamiltonian. The quality of this approximation depends on the truncation criterion, which we will comment on later, and the truncation dimension $\chi$. The errors incurred by the truncation can be quantified by comparing the TGBA data (e.g., physical observables) with successively increasing $\chi$.

If the Hamiltonian $H$ has a symmetry which is shared by $H_0$ and $V$, it can be fully exploited in the TGBA. The truncated Hilbert space has a block structure, $\mathcal{H}_{\mathrm{trunc}} = \oplus_{a} \mathcal{H}_{a,\mathrm{trunc}}$, where $a$ labels different symmetry sectors. Using the fermionic Gaussian eigenbasis of $H_0$, $|\phi_{a,\alpha}\rangle \in \mathcal{H}_{a,\mathrm{trunc}}$, a truncated Hamiltonian is defined in each sector, $H_{(a,\alpha);(a,\beta)} = \langle \phi_{a,\alpha}|H|\phi_{a,\beta} \rangle$, which can be diagonalized independently. This allows one to carry out parallel computation and simulate much larger system sizes which would be impossible otherwise. In the Ising chain example below, we will illustrate how to make use of translation symmetry in the TGBA.

The truncation criterion determines whether the TGBA will succeed or not. For most applications in equilibrium (e.g., determining the spectrum of $H$ or calculating dynamic correlation functions), a natural way to perform the truncation is to retain $\chi$ lowest energy states of $H_0$ in $\mathcal{H}_\mathrm{trunc}$. This is of the same spirit as Wilson's NRG~\cite{Wilson1975} and the truncated conformal space approach~\cite{Yurov1990,Yurov1991,Lassig1991,Beria2013,Konik2015,James2018,Mussardo-Book} and can be justified when the $V$ term is weak in some sense. However, for certain applications, different truncation schemes might be more suited to construct $\mathcal{H}_\mathrm{trunc}$. For instance, when simulating quench dynamics, we observe that retaining those states which have largest overlaps with the initial state is a more appropriate metric, as we shall discuss below.

{\em Example} --- To illustrate how the TGBA works, we choose the quantum Ising chain with both transverse and longitudinal fields as a paradigmatic example:
\begin{align}
H_0 &= -\sum_{j=1}^N \sigma^x_j\sigma^x_{j+1} -  h_\perp \sum_{j=1}^N\sigma^z_j \, , \nonumber \\
V &= - h_\parallel \sum_{j=1}^N\sigma^x_j \, ,
\label{eq:Hamil}
\end{align}
where $\sigma^{\alpha}_j$ ($\alpha=x,y,z$) are Pauli operators at site $j$ and $N$ is the total number of sites. We consider $h_{\perp} >0$ and even $N$ throughout this work and impose periodic boundary conditions ($\sigma^{\alpha}_{N+1} = \sigma^{\alpha}_1$) to preserve translation symmetry.

As the ``free part'' of the full Hamiltonian, $H_0$ corresponds to the transverse-field Ising chain~\cite{Pfeuty1970}, which, using the Jordan-Wigner transformation $\sigma_j^x =  (c_j+c_j^\dagger) (-1)^{\sum_{l=1}^{j-1}
c_l^{\dagger} c_l}$ and $\sigma_j^z = 2c_j^{\dagger} c_j-1$, is mapped to a fermionic Hamiltonian:
\begin{align}
    H_0 &= \sum_{j=1}^{N-1}(c_j-c_j^\dagger)(c_{j+1}+c_{j+1}^\dagger)- Q (c_N-c_N^\dagger)(c_1+c_1^\dagger)\nonumber\\
    &\phantom{=} - h_\perp \sum_{j=1}^N (2c_j^\dagger c_j-1) \, .
\label{eq:H0-fermion}
\end{align}
The fermion parity operator $Q =(-1)^{\sum_{j=1}^N c_j^\dagger c_j}$ commutes with $H_0$ and has eigenvalues $\pm 1$. Thus, the eigenstates of $H_0$ fall into two sectors: the Neveu-Schwarz (NS) sector with $Q=1$, and the Ramond (R) sector with $Q=-1$. In the NS (R) sector, $H_0$ in Eq.~\eqref{eq:H0-fermion} becomes quadratic by imposing anti-periodic (periodic) boundary condition $c_{N+1} = -c_1$ ($c_{N+1} = c_1$) for the Jordan-Wigner fermions.
After performing the Fourier transformation $c_j =\frac{1}{\sqrt{N}}\sum_{k\in \mathrm{NS/R}} e^{ikj} c_k$ and a Bogouliubov transformation, $H_0$ is diagonalized as
\begin{align}
    H_0^{\mathrm{NS/R}} = \sum_{k\in\mathrm{NS/R}} \varepsilon_k \left(d_k^\dagger d_k -\frac{1}{2}\right)
\label{eq:H0-diagonalized}
\end{align}
with
\begin{align}
    \varepsilon_k &=\begin{cases}
    2\sqrt{(h_\perp-1)^2+4h_\perp\cos^2(k/2)} &\quad k \neq\pi  \\
    2(h_\perp-1)  &\quad k=\pi \\
    \end{cases},
\end{align}
where allowed single-particle momenta in the NS and R sector are $k = \pm \frac{\pi}{N}, \pm \frac{3\pi}{N},\ldots, \pm \frac{(N-1)\pi}{N}$ and $k = 0, \pm \frac{2\pi}{N}, \ldots, \pm \frac{(N-2)\pi}{N}, \pi$, respectively. The Bogoliubov mode in Eq.~\eqref{eq:H0-diagonalized} is given by $d_k = \sin(\theta_k/2)c_k -i\cos(\theta_k/2)c_{-k}^\dagger$ with $\theta_k = \mathrm{sign}(k) \, \arccos\frac{2(h_\perp + \cos k)}{\varepsilon_k} \in (-\pi,\pi]$. The ground state in the NS (R) sector reads $|0\rangle_{\mathrm{NS}} = [\prod_{k>0}\cos(\theta_k/2)]^{-1}\prod_{k\in\mathrm{NS}}d_k |0\rangle_c$ ($|0\rangle_{\mathrm{R}} = [\prod_{k>0}\cos(\theta_k/2)]^{-1} \prod_{k\in\mathrm{R},k\neq \pi}d_k |0\rangle_c$), where $|0\rangle_c$ is the vacuum of Jordan-Wigner fermions. Thus, all eigenstates of $H_0$ are written as $d^\dagger_{k_1} d^\dagger_{k_2} \cdots d^\dagger_{k_{M}} |0\rangle_{\mathrm{NS/R}}$  ($M$ even), with distinct single-particle momenta $k_1<k_2<\cdots < k_M$ in the respective sector. They form the full Hilbert space $\mathcal{H} = \oplus_k \mathcal{H}_k$, where the many-body momentum of eigenstates, $k = \sum_{j=1}^M k_j \, (\mathrm{mod} \; 2\pi)$, labels different symmetry sectors.

To set up the TGBA, we define the truncated subspace as $\mathcal{H}_{\mathrm{trunc}} = \oplus_{k} \mathcal{H}_{k,\mathrm{trunc}}$, with basis vectors $\{|\phi_{k,\alpha}\rangle\}$ ($\alpha = 1,\ldots,\chi$)~\footnote{For simplicity, we use the same truncation dimension $\chi$ across all momentum sectors.}. The retained basis vectors are selected from the eigenstates of $H_0$ described above, according to which selection criterion depends on the specific application. A crucial technical step in the TGBA is to compute the matrix elements of the $V$ term [Eq.~\eqref{eq:Hamil}] in the truncated subspace. As the $V$ term is translation-invariant, it does not mix basis vectors with different momenta. Within the $\mathcal{H}_{k,\mathrm{trunc}}$ subspace, matrix elements of $V$ read $\langle \phi_{k,\alpha} |V|\phi_{k,\beta}\rangle = -N h_\parallel \langle \phi_{k, \alpha} |(c_1 + c_1^\dagger)|\phi_{k, \beta}\rangle $. Using Wick's theorem~\cite{Append}, evaluating such matrix elements reduces to calculating the Pfaffian of a matrix~\cite{Bertsch2012,Carlsson2021,JinHK2022,Mascot2023} with dimension $\mathcal{O}(N)$, this allowed us to set up the effective Hamiltonian efficiently for system sizes up to $N \sim 100$~\cite{Wimmer2012}.

{\em Dynamic structure factor} --- We first demonstrate the computation of the longitudinal dynamical structure factor
\begin{align}
    S^{xx}(k, \omega) = \frac{1}{N} \sum_{j,l=1}^N e^{-ik(j-l)} \int_{-\infty}^\infty \mathrm{d}t \, e^{i \omega t} \langle \sigma_j^x(t) \sigma_{l}^x(0) \rangle
\label{eq:Sxx-dynamic}
\end{align}
for the Ising chain \eqref{eq:Hamil} at zero temperature. For this calculation, we adopt $h_\perp = 1$ and $h_\parallel = 0.05$ with chain length $N=60$. The model can hence be viewed as a critical Ising chain perturbed by a longitudinal field, whose low-energy effective theory is the integrable $E_8$ Toda field theory exhibiting eight massive excitations with universal mass ratios~\cite{Zamolodchikov1989}.

For computing the dynamic structure factor, we retain $\chi = 1600$ lowest energy eigenstates of $H_0$ in $\mathcal{H}_{k,\mathrm{trunc}}$ and diagonalize the effective Hamiltonians in each momentum sector to obtain (approximate) eigenstates $|\psi_{k,\alpha} \rangle = \sum_{\beta=1}^{\chi} [M(k)]_{\alpha \beta} |\phi_{k,\beta} \rangle$ as well as their corresponding energies. The ground state lies in the $k=0$ sector, which we denote as $|\psi_{0,1}\rangle$ (with energy $E_{0,1}$). Inserting the approximated time evolution operator $e^{-iHt} \simeq \sum_k \sum_{\alpha=1}^{\chi} e^{-iE_{k,\alpha}t} |\psi_{k,\alpha}\rangle \langle \psi_{k,\alpha}|$ into Eq.~\eqref{eq:Sxx-dynamic}, the dynamical structure factor becomes
\begin{align}
    S^{xx}(k, \omega) \simeq 2 \pi \sum_{\alpha=1}^{\chi} |\langle \psi_{0,1}| \sigma_k^x |\psi_{k,\alpha} \rangle |^2 \delta(\omega + E_{0,1} - E_{k,\alpha})
\label{eq:dsf}
\end{align}
with $\sigma_k^x = \frac{1}{\sqrt{N}}\sum_{j=1}^N e^{-ik j} \sigma_j^x $. Using translation symmetry, the matrix elements in Eq.~\eqref{eq:dsf}, $\langle \psi_{0,1}| \sigma_k^x |\psi_{k,\alpha} \rangle = \sqrt{N} e^{-ik} \sum_{\mu,\nu=1}^{\chi} [M(0)]^{*}_{1,\mu}\langle \phi_{0,\mu}|(c^{\dag}_1 + c_1)| \phi_{k,\nu} \rangle [M(k)]_{\alpha,\nu}$, also reduce to Pfaffian computations~\cite{Append}.

\begin{figure}
\resizebox{\columnwidth}{!}{\includegraphics{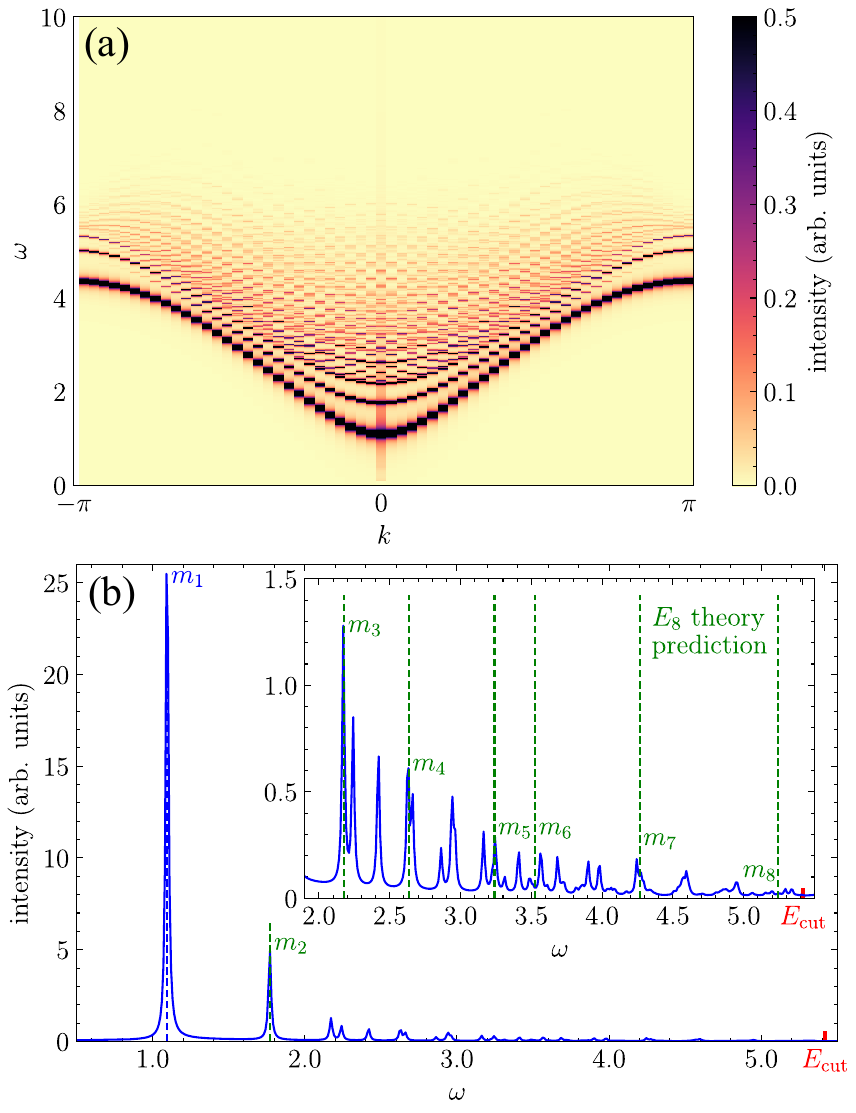}}
\caption{(a) Dynamic structure factor $S^{xx}(k, \omega)$ at $h_\perp=1$ and $h_\parallel=0.05$ for a chain of $N=60$ sites. (b) Dynamic structure factor $S^{xx}(k=0, \omega)$. The dashed lines represent the analytical predictions of the $E_8$ masses.
}
\label{fig:E8}
\end{figure}

The numerically computed $S^{xx}(k,\omega)$ is depicted in Fig.~\ref{fig:E8}(a). It provides spectral information beyond the low-energy limit and is readily comparable to experiments. The $E_8$ Toda field theory predicts eight massive excitations~\cite{Zamolodchikov1989,Kjaell2011}, whose masses are universal up to a mass scale set by $m_1 \sim h_\parallel^{8/15}$~\footnote{To ensure the results are not dominated by finite-size effects, the correlation length $\xi \sim m_1^{-1} \sim h_\parallel^{-8/15}$ should be small compared to the length of the chain. We find that for $h_\parallel = 0.05$ this is still the case.}. Seven out of the eight massive particles can be reliably identified in $S^{xx}(k=0,\omega)$, as shown in Fig.~\ref{fig:E8}(b). The retained $\chi = 1600$ states in $k=0$ sector corresponds to a cutoff energy at $E_\mathrm{cut} \simeq 5.4 \simeq 4.9 \, m_1$. The mass of the heaviest particle (analytical value: $m_8 \simeq 4.78 \, m_1$) is too close to $E_\mathrm{cut}$ and has not converged properly in our $\chi = 1600$ calculation. For comparison, the numerical values of six mass ratios from the TGBA calculation and the values predicted by the $E_8$ Toda field theory are displayed in Table~\ref{tab:E8-ratios}.

\begin{table}[]
    \resizebox{\columnwidth}{!}{
    \begin{tabular}{c|llllll}
        & $m_2/m_1$ & $m_3/m_1$ & $m_4/m_1$ & $m_5/m_1$ & $m_6/m_1$ & $m_7/m_1$ \\
        \hline
        TGBA & 1.616 & 1.982 & 2.401 & 2.959 & 3.260 & 3.887 \\
        $E_8$ theory & 1.618 & 1.989 & 2.405 & 2.956 & 3.218 & 3.891
    \end{tabular}
    }
    \caption{Peak ratios of the $E_8$ single-particle states obtained from the TGBA calculation versus the analytical prediction from the $E_8$ Toda field theory~\cite{Zamolodchikov1989}.}
    \label{tab:E8-ratios}
\end{table}

{\em Quench dynamics} --- We now employ the TGBA to simulate out of equilibrium dynamics following a quantum quench. Hamiltonian truncation methods have already been used to simulate quench dynamics in the Ising field theory, using either the Ising conformal field theory or the massive Majorana field theory as a basis to construct the truncated subspace \cite{Rakovszky2016, Kukuljan2018, Hodsagi2018}.
The method we employ here is of a similar spirit, however we found that choosing a different truncation scheme can improve the performance of the TGBA significantly.

We consider the following global quench protocol: The system is prepared in some initial state $|\Phi_0\rangle$, which we choose to be an eigenstate of the transverse-field Ising chain $H_0(t=0)$. Then both the transverse and longitudinal fields are quenched. To perform the simulation, we decompose the Hamiltonian after the quench again into a ``free'' and an interacting part, $H(t>0) = H_0(t>0) + V(t>0)$, and apply the TGBA to approximate the time evolution operator $e^{-iH(t>0)t} \simeq \sum_{\alpha=1}^{\chi} e^{-iE_\alpha t} |\psi_\alpha \rangle \langle \psi_\alpha|$ using the eigenbasis of $H_0(t>0)$. Denoting this eigenbasis again by $\{|\phi_\alpha\rangle \}$, we write the time-evolved state as
\begin{align}
    e^{-iH(t>0)t}|\Phi_0\rangle \simeq  \sum_{\alpha=1}^\chi B_\alpha(t) |\phi_\alpha \rangle
\label{eq:time-evolved-expansion}
\end{align}
with $B_{\alpha}(t) = \sum_{\beta, \gamma=1}^{\chi} M^T_{ \alpha \beta} e^{-iE_{\beta}t} M_{\beta \gamma}^* \langle\phi_{\gamma} | \Phi_0 \rangle$.
These coefficients contain not only the approximated eigenenergies $E_\beta$ and the variational coefficients $M_{\alpha \beta}$ obtained from the TGBA, but also the overlaps $\langle \phi_{\alpha}|\Phi_0\rangle$ of the initial state with states spanning the truncated subspace. Hence, to fully capture the dynamics of the time-evolved state, it is not only necessary to obtain a good approximation for the spectrum and eigenstates of $H(t>0)$, one must also make sure that the truncated subspace adequately captures the initial state $|\Phi_0\rangle$. Accordingly, we construct the truncated subspace by ordering the states by the absolute value of their overlap with $|\Phi_0\rangle$ instead of their bare energies $E^{(0)}_\alpha$. How well the initial state is captured by the truncated subspace can be conveniently checked by calculating $\sum_{\alpha=1}^\chi |\langle \phi_\alpha|\Phi_0\rangle |^2 \leq \langle \Phi_0|\Phi_0 \rangle = 1$. It should also be noted that the TGBA approximation for the time-evolved state is not exactly normalized, but its norm is time-independent and coincides with the overlap of the truncated subspace basis states with the initial state, $|| e^{-iH(t>0)t} |\Phi_0\rangle || \simeq \sqrt{\sum_{\alpha = 1}^\chi |\langle \phi_\alpha|\Phi_0\rangle |^2}$.

Rather than scanning the entire Gaussian basis $\{ |\phi_{\alpha}\rangle \}$, we adopt a more practical approach by first truncating based on the number of Bogoliubov quasiparticles~\cite{Append}. The full Gaussian basis $\{ |\phi_{\alpha}\rangle \}$ consists of the Bogoliubov vacuum and ``excited'' states with a certain number of Bogoliubov quasiparticles on top of the Bogoliubov vacuum. Our algorithm proceeds as follows: (i) Set a maximum number of Bogoliubov quasiparticles for constructing $\mathcal{H}_{\mathrm{trunc}}$; (ii) For each allowed quasiparticle number, generate all corresponding basis states, sort them by the overlap $|\langle \phi_\alpha|\Phi_0\rangle|$, and retain those with overlap above a chosen threshold. In our simulations, we typically set the maximum number of quasiparticles to be in the range of $8$ to $10$.

To obtain a quantitative and observable-independent measure of the convergence of the TGBA data, we compute the norm of the difference between two successive TGBA approximations with different truncation dimensions. Since this quantity is time-dependent, we define a metric between approximations by taking the maximum of this difference over some time span $[0, t_\text{max}]$. In terms of the coefficients $B_\alpha(t)$, this metric is given by
\begin{align}
    & \phantom{=} \quad \varepsilon(\chi^\prime, \chi)  \nonumber \\
    &= \max_{t \in [0, t_\text{max}]} || |\Psi^\chi(t)\rangle - |\Psi^{\chi^\prime}(t)\rangle || \nonumber \\
    &= \max_{t \in [0, t_\text{max}]} \sqrt{\sum_{\alpha=1}^\chi |B_\alpha^\chi(t) - B_\alpha^{\chi^\prime}(t)|^2 + \sum_{\alpha=\chi+1}^{\chi^\prime} |B_\alpha^{\chi^\prime}(t)|^2},
\label{eq:quench_error}
\end{align}
where $|\Psi^\chi(t)\rangle$ and $|\Psi^{\chi^\prime}(t)\rangle$ denote two approximations of the time-evolved state with truncation dimension $\chi$ and $\chi^\prime$ ($\chi < \chi^\prime$), respectively. $B^\chi_\alpha(t)$ and $ B^{\chi^\prime}_\alpha(t)$ are the corresponding coefficients in the respective truncated subspaces.

\begin{figure}
    \resizebox{\columnwidth}{!}{\includegraphics{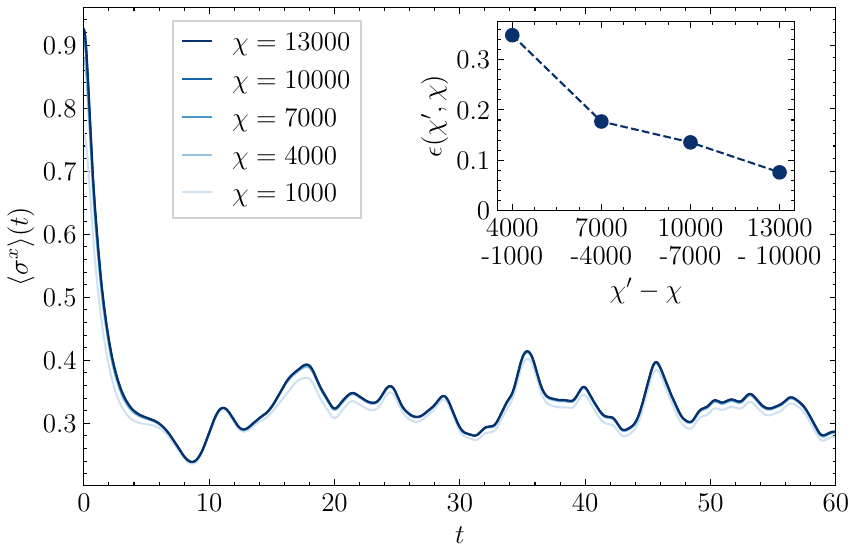}}
\caption{Longitudinal magnetization following a global quench $(h_\perp, h_\parallel) = (0.5, 0) \mapsto (1.0, 0.05)$ from the initial state $|\Phi_0\rangle = \frac{1}{\sqrt{2}} (|0\rangle_\mathrm{NS} \pm |0\rangle_\mathrm{R})$ with positive magnetization, for a chain of $N=60$ sites. (a) Time evolution of the longitudinal magnetization $\langle \sigma^x \rangle(t)$ up to $t_\text{max} = 60$. Inset: metric $\varepsilon$ between successive approximations, as defined in Eq.~\eqref{eq:quench_error}. }
\label{fig:order_parameter}
\end{figure}

To test the method, we calculate the time evolution of the longitudinal magnetization $\langle \sigma^x \rangle (t)  \equiv \frac{1}{N} \sum_{j=1}^N\langle \sigma^x_j \rangle(t)$ under the following quench protocol: at $t=0$, the system is initialized in the ferromagnetic phase of the transverse-field Ising chain with $\langle \sigma^x \rangle > 0$, for some $h_\perp(t=0) < 1$. For $t>0$, the Hamiltonian parameters suddenly change to those in the $E_8$ region, namely $h_\perp(t>0) = 1$ and $h_\parallel(t>0) > 0$. Since the initial state has many-body momentum $k=0$ and the quench Hamiltonian preserves translation symmetry, only states from the $k=0$ sector need to be included in the truncated subspace. Figure~\ref{fig:order_parameter} shows the time evolution of the order parameter, $\langle \sigma^x \rangle(t)$, following a quench $(h_\perp, h_\parallel) = (0.5, 0) \mapsto (1.0, 0.05)$ with different truncation dimensions in the TGBA. These results, along with the Fourier transform (see Appendix~\cite{Append}), demonstrate good convergence with increasing $\chi$.

\begin{figure*}
    \resizebox{2\columnwidth}{!}{\includegraphics{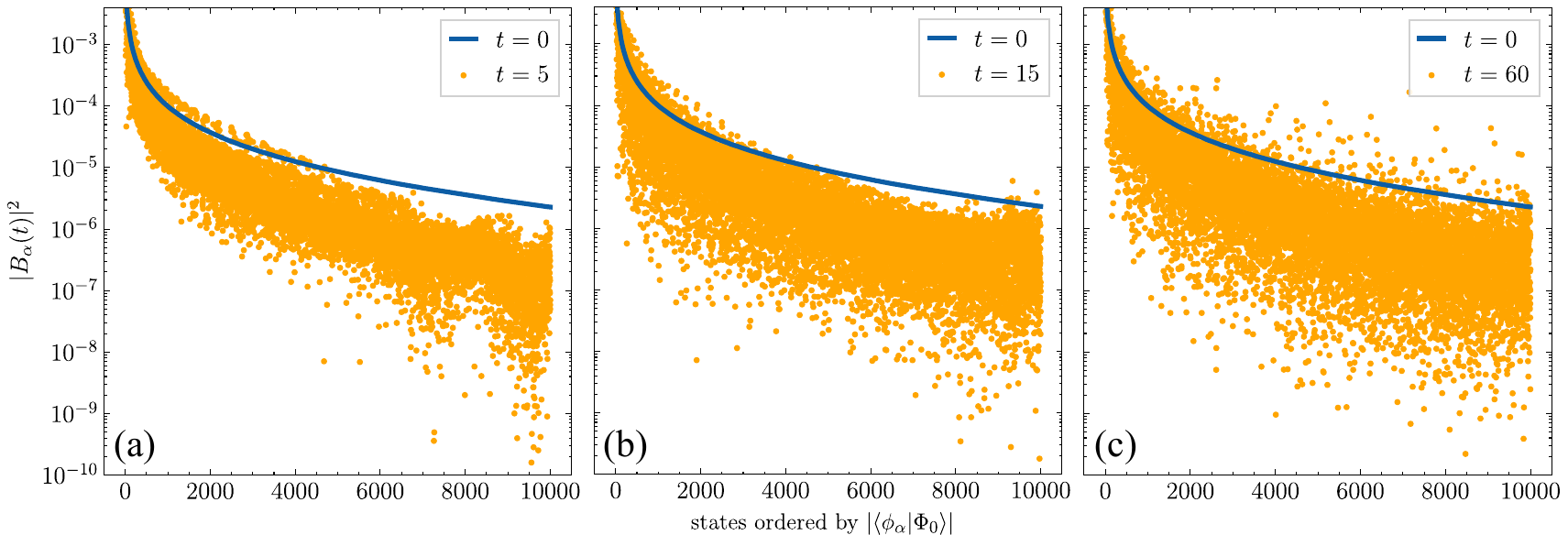}}
\caption{Time evolution of the squared amplitude $|B_{\alpha}(t)|^2$ [see Eq.~\eqref{eq:time-evolved-expansion}] of the time-evolved state expanded in the fermionic Gaussian basis. The quench protocol is $(h_\perp, h_\parallel) = (0.5, 0) \mapsto (1.0, 0.05)$, and the system size is $N=60$. Blue lines indicate the initial amplitudes at $t=0$. Panels show snapshots at (a) $t=5$, (b) $t=15$, and (c) $t=60$.}
\label{fig:coefficients}
\end{figure*}

As mentioned earlier, the success of the quench dynamics simulation is closely tied to the truncation criterion. For small values of $h_\parallel$, we observe that the TGBA remains convergent up to very long times, provided that the initial state is sufficiently well captured within the truncated subspace. In Fig.~\ref{fig:coefficients}, we show that fermionic Gaussian states with initially large overlaps continue to dominate and thus should be retained in $\mathcal{H}_{\mathrm{trunc}}$, and only a small number of states that were initially less important gain appreciable weight as time evolves.

{\em Summary and outlook} --- We have proposed a new method, the TGBA, to simulate dynamics of quantum many-body systems. Using the non-integrable quantum Ising chain as a benchmark, we demonstrate that, with appropriate truncation schemes, the TGBA can accurately reproduce the universal mass ratios predicted by the $E_8$ field theory and capture spectral features beyond the low-energy limit with full momentum resolution. A key advantage of the TGBA is that the effective Hamiltonian can be efficiently constructed and simulated within the truncated Gaussian subspace, enabling simulations both in and out of equilibrium.

For the cases considered here, convergence was typically achieved for truncation dimensions of order $\chi \sim 1000 - 10000$, a regime where the computational cost of diagonalizing the effective Hamiltonian remains negligible. For more demanding applications, it may be necessary to adopt an NRG-like scheme~\cite{Konik2007} to construct the truncated basis iteratively, thereby allowing access to much larger truncation dimensions. It will also be interesting to apply the TGBA to other interesting physical systems, such as the Hubbard model with small or moderate $U$.

{\em Acknowledgments} --- We are grateful to Mari Carmen Ba\~nuls, Jan Carl Budich, Jan von Delft, Andreas Gleis, Michael Knap, Oleksandra Kovalska, Goran Nakerst, Frank Pollmann, Wei Tang, Lei Wang, Tao Xiang, and Haiyuan Zou for helpful discussions. N.A. is supported by the Deutsche Forschungsgemeinschaft (DFG) through project A06 of SFB 1143 (Project No.~247310070). Y.S.Z. is supported by the Sino-German (CSC-DAAD) Postdoc Scholarship Program. The authors gratefully acknowledge the computing time made available to them on the high-performance computer at the NHR Center of TU
Dresden. This center is jointly supported by the Federal Ministry of Education and Research and the state governments participating in the NHR~\footnote{\url{www.nhr-verein.de/unsere-partner}}.

\bibliography{refs}

\begin{thebibliography}{62}%
\makeatletter
\providecommand \@ifxundefined [1]{%
 \@ifx{#1\undefined}
}%
\providecommand \@ifnum [1]{%
 \ifnum #1\expandafter \@firstoftwo
 \else \expandafter \@secondoftwo
 \fi
}%
\providecommand \@ifx [1]{%
 \ifx #1\expandafter \@firstoftwo
 \else \expandafter \@secondoftwo
 \fi
}%
\providecommand \natexlab [1]{#1}%
\providecommand \enquote  [1]{``#1''}%
\providecommand \bibnamefont  [1]{#1}%
\providecommand \bibfnamefont [1]{#1}%
\providecommand \citenamefont [1]{#1}%
\providecommand \href@noop [0]{\@secondoftwo}%
\providecommand \href [0]{\begingroup \@sanitize@url \@href}%
\providecommand \@href[1]{\@@startlink{#1}\@@href}%
\providecommand \@@href[1]{\endgroup#1\@@endlink}%
\providecommand \@sanitize@url [0]{\catcode `\\12\catcode `\$12\catcode
  `\&12\catcode `\#12\catcode `\^12\catcode `\_12\catcode `\%12\relax}%
\providecommand \@@startlink[1]{}%
\providecommand \@@endlink[0]{}%
\providecommand \url  [0]{\begingroup\@sanitize@url \@url }%
\providecommand \@url [1]{\endgroup\@href {#1}{\urlprefix }}%
\providecommand \urlprefix  [0]{URL }%
\providecommand \Eprint [0]{\href }%
\providecommand \doibase [0]{http://dx.doi.org/}%
\providecommand \selectlanguage [0]{\@gobble}%
\providecommand \bibinfo  [0]{\@secondoftwo}%
\providecommand \bibfield  [0]{\@secondoftwo}%
\providecommand \translation [1]{[#1]}%
\providecommand \BibitemOpen [0]{}%
\providecommand \bibitemStop [0]{}%
\providecommand \bibitemNoStop [0]{.\EOS\space}%
\providecommand \EOS [0]{\spacefactor3000\relax}%
\providecommand \BibitemShut  [1]{\csname bibitem#1\endcsname}%
\let\auto@bib@innerbib\@empty
\bibitem [{\citenamefont {Polkovnikov}\ \emph {et~al.}(2011)\citenamefont
  {Polkovnikov}, \citenamefont {Sengupta}, \citenamefont {Silva},\ and\
  \citenamefont {Vengalattore}}]{Plokovnikov2011}%
  \BibitemOpen
  \bibfield  {author} {\bibinfo {author} {\bibfnamefont {A.}~\bibnamefont
  {Polkovnikov}}, \bibinfo {author} {\bibfnamefont {K.}~\bibnamefont
  {Sengupta}}, \bibinfo {author} {\bibfnamefont {A.}~\bibnamefont {Silva}}, \
  and\ \bibinfo {author} {\bibfnamefont {M.}~\bibnamefont {Vengalattore}},\
  }\href {\doibase 10.1103/RevModPhys.83.863} {\bibfield  {journal} {\bibinfo
  {journal} {Rev. Mod. Phys.}\ }\textbf {\bibinfo {volume} {83}},\ \bibinfo
  {pages} {863} (\bibinfo {year} {2011})}\BibitemShut {NoStop}%
\bibitem [{\citenamefont {Georgescu}\ \emph {et~al.}(2014)\citenamefont
  {Georgescu}, \citenamefont {Ashhab},\ and\ \citenamefont
  {Nori}}]{Georgescu2014}%
  \BibitemOpen
  \bibfield  {author} {\bibinfo {author} {\bibfnamefont {I.~M.}\ \bibnamefont
  {Georgescu}}, \bibinfo {author} {\bibfnamefont {S.}~\bibnamefont {Ashhab}}, \
  and\ \bibinfo {author} {\bibfnamefont {F.}~\bibnamefont {Nori}},\ }\href
  {\doibase 10.1103/RevModPhys.86.153} {\bibfield  {journal} {\bibinfo
  {journal} {Rev. Mod. Phys.}\ }\textbf {\bibinfo {volume} {86}},\ \bibinfo
  {pages} {153} (\bibinfo {year} {2014})}\BibitemShut {NoStop}%
\bibitem [{\citenamefont {D’Alessio}\ \emph {et~al.}(2016)\citenamefont
  {D’Alessio}, \citenamefont {Kafri}, \citenamefont {Polkovnikov},\ and\
  \citenamefont {Rigol}}]{DAlessio2016}%
  \BibitemOpen
  \bibfield  {author} {\bibinfo {author} {\bibfnamefont {L.}~\bibnamefont
  {D’Alessio}}, \bibinfo {author} {\bibfnamefont {Y.}~\bibnamefont {Kafri}},
  \bibinfo {author} {\bibfnamefont {A.}~\bibnamefont {Polkovnikov}}, \ and\
  \bibinfo {author} {\bibfnamefont {M.}~\bibnamefont {Rigol}},\ }\href
  {\doibase 10.1080/00018732.2016.1198134} {\bibfield  {journal} {\bibinfo
  {journal} {Adv. Phys.}\ }\textbf {\bibinfo {volume} {65}},\ \bibinfo {pages}
  {239} (\bibinfo {year} {2016})}\BibitemShut {NoStop}%
\bibitem [{\citenamefont {Abanin}\ \emph {et~al.}(2019)\citenamefont {Abanin},
  \citenamefont {Altman}, \citenamefont {Bloch},\ and\ \citenamefont
  {Serbyn}}]{Abanin2019}%
  \BibitemOpen
  \bibfield  {author} {\bibinfo {author} {\bibfnamefont {D.~A.}\ \bibnamefont
  {Abanin}}, \bibinfo {author} {\bibfnamefont {E.}~\bibnamefont {Altman}},
  \bibinfo {author} {\bibfnamefont {I.}~\bibnamefont {Bloch}}, \ and\ \bibinfo
  {author} {\bibfnamefont {M.}~\bibnamefont {Serbyn}},\ }\href {\doibase
  10.1103/RevModPhys.91.021001} {\bibfield  {journal} {\bibinfo  {journal}
  {Rev. Mod. Phys.}\ }\textbf {\bibinfo {volume} {91}},\ \bibinfo {pages}
  {021001} (\bibinfo {year} {2019})}\BibitemShut {NoStop}%
\bibitem [{\citenamefont {Vidal}(2003)}]{Vidal2003}%
  \BibitemOpen
  \bibfield  {author} {\bibinfo {author} {\bibfnamefont {G.}~\bibnamefont
  {Vidal}},\ }\href {\doibase 10.1103/PhysRevLett.91.147902} {\bibfield
  {journal} {\bibinfo  {journal} {Phys. Rev. Lett.}\ }\textbf {\bibinfo
  {volume} {91}},\ \bibinfo {pages} {147902} (\bibinfo {year}
  {2003})}\BibitemShut {NoStop}%
\bibitem [{\citenamefont {White}\ and\ \citenamefont
  {Feiguin}(2004)}]{White2003}%
  \BibitemOpen
  \bibfield  {author} {\bibinfo {author} {\bibfnamefont {S.~R.}\ \bibnamefont
  {White}}\ and\ \bibinfo {author} {\bibfnamefont {A.~E.}\ \bibnamefont
  {Feiguin}},\ }\href {\doibase 10.1103/PhysRevLett.93.076401} {\bibfield
  {journal} {\bibinfo  {journal} {Phys. Rev. Lett.}\ }\textbf {\bibinfo
  {volume} {93}},\ \bibinfo {pages} {076401} (\bibinfo {year}
  {2004})}\BibitemShut {NoStop}%
\bibitem [{\citenamefont {Daley}\ \emph {et~al.}(2004)\citenamefont {Daley},
  \citenamefont {Kollath}, \citenamefont {Schollwöck},\ and\ \citenamefont
  {Vidal}}]{Daley2004}%
  \BibitemOpen
  \bibfield  {author} {\bibinfo {author} {\bibfnamefont {A.~J.}\ \bibnamefont
  {Daley}}, \bibinfo {author} {\bibfnamefont {C.}~\bibnamefont {Kollath}},
  \bibinfo {author} {\bibfnamefont {U.}~\bibnamefont {Schollwöck}}, \ and\
  \bibinfo {author} {\bibfnamefont {G.}~\bibnamefont {Vidal}},\ }\href
  {\doibase 10.1088/1742-5468/2004/04/p04005} {\bibfield  {journal} {\bibinfo
  {journal} {J. Stat. Mech.: Theory Exp.}\ }\textbf {\bibinfo {volume}
  {2004}},\ \bibinfo {pages} {P04005} (\bibinfo {year} {2004})}\BibitemShut
  {NoStop}%
\bibitem [{\citenamefont {Haegeman}\ \emph {et~al.}(2011)\citenamefont
  {Haegeman}, \citenamefont {Cirac}, \citenamefont {Osborne}, \citenamefont
  {Pi\ifmmode~\check{z}\else \v{z}\fi{}orn}, \citenamefont {Verschelde},\ and\
  \citenamefont {Verstraete}}]{Haegeman2011}%
  \BibitemOpen
  \bibfield  {author} {\bibinfo {author} {\bibfnamefont {J.}~\bibnamefont
  {Haegeman}}, \bibinfo {author} {\bibfnamefont {J.~I.}\ \bibnamefont {Cirac}},
  \bibinfo {author} {\bibfnamefont {T.~J.}\ \bibnamefont {Osborne}}, \bibinfo
  {author} {\bibfnamefont {I.}~\bibnamefont {Pi\ifmmode~\check{z}\else
  \v{z}\fi{}orn}}, \bibinfo {author} {\bibfnamefont {H.}~\bibnamefont
  {Verschelde}}, \ and\ \bibinfo {author} {\bibfnamefont {F.}~\bibnamefont
  {Verstraete}},\ }\href {\doibase 10.1103/PhysRevLett.107.070601} {\bibfield
  {journal} {\bibinfo  {journal} {Phys. Rev. Lett.}\ }\textbf {\bibinfo
  {volume} {107}},\ \bibinfo {pages} {070601} (\bibinfo {year}
  {2011})}\BibitemShut {NoStop}%
\bibitem [{\citenamefont {Haegeman}\ \emph {et~al.}(2016)\citenamefont
  {Haegeman}, \citenamefont {Lubich}, \citenamefont {Oseledets}, \citenamefont
  {Vandereycken},\ and\ \citenamefont {Verstraete}}]{Haegeman2016}%
  \BibitemOpen
  \bibfield  {author} {\bibinfo {author} {\bibfnamefont {J.}~\bibnamefont
  {Haegeman}}, \bibinfo {author} {\bibfnamefont {C.}~\bibnamefont {Lubich}},
  \bibinfo {author} {\bibfnamefont {I.}~\bibnamefont {Oseledets}}, \bibinfo
  {author} {\bibfnamefont {B.}~\bibnamefont {Vandereycken}}, \ and\ \bibinfo
  {author} {\bibfnamefont {F.}~\bibnamefont {Verstraete}},\ }\href {\doibase
  10.1103/PhysRevB.94.165116} {\bibfield  {journal} {\bibinfo  {journal} {Phys.
  Rev. B}\ }\textbf {\bibinfo {volume} {94}},\ \bibinfo {pages} {165116}
  (\bibinfo {year} {2016})}\BibitemShut {NoStop}%
\bibitem [{\citenamefont {Li}\ \emph {et~al.}(2024)\citenamefont {Li},
  \citenamefont {Gleis},\ and\ \citenamefont {von Delft}}]{LiJW2024}%
  \BibitemOpen
  \bibfield  {author} {\bibinfo {author} {\bibfnamefont {J.-W.}\ \bibnamefont
  {Li}}, \bibinfo {author} {\bibfnamefont {A.}~\bibnamefont {Gleis}}, \ and\
  \bibinfo {author} {\bibfnamefont {J.}~\bibnamefont {von Delft}},\ }\href
  {\doibase 10.1103/PhysRevLett.133.026401} {\bibfield  {journal} {\bibinfo
  {journal} {Phys. Rev. Lett.}\ }\textbf {\bibinfo {volume} {133}},\ \bibinfo
  {pages} {026401} (\bibinfo {year} {2024})}\BibitemShut {NoStop}%
\bibitem [{\citenamefont {Jarrell}\ and\ \citenamefont
  {Gubernatis}(1996)}]{Jarrell1996}%
  \BibitemOpen
  \bibfield  {author} {\bibinfo {author} {\bibfnamefont {M.}~\bibnamefont
  {Jarrell}}\ and\ \bibinfo {author} {\bibfnamefont {J.}~\bibnamefont
  {Gubernatis}},\ }\href {\doibase
  https://doi.org/10.1016/0370-1573(95)00074-7} {\bibfield  {journal} {\bibinfo
   {journal} {Phys. Rep.}\ }\textbf {\bibinfo {volume} {269}},\ \bibinfo
  {pages} {133} (\bibinfo {year} {1996})}\BibitemShut {NoStop}%
\bibitem [{\citenamefont {Sandvik}(1998)}]{Sandvik1998}%
  \BibitemOpen
  \bibfield  {author} {\bibinfo {author} {\bibfnamefont {A.~W.}\ \bibnamefont
  {Sandvik}},\ }\href {\doibase 10.1103/PhysRevB.57.10287} {\bibfield
  {journal} {\bibinfo  {journal} {Phys. Rev. B}\ }\textbf {\bibinfo {volume}
  {57}},\ \bibinfo {pages} {10287} (\bibinfo {year} {1998})}\BibitemShut
  {NoStop}%
\bibitem [{\citenamefont {Fei}\ \emph {et~al.}(2021)\citenamefont {Fei},
  \citenamefont {Yeh},\ and\ \citenamefont {Gull}}]{Fei2021}%
  \BibitemOpen
  \bibfield  {author} {\bibinfo {author} {\bibfnamefont {J.}~\bibnamefont
  {Fei}}, \bibinfo {author} {\bibfnamefont {C.-N.}\ \bibnamefont {Yeh}}, \ and\
  \bibinfo {author} {\bibfnamefont {E.}~\bibnamefont {Gull}},\ }\href
  {http://dx.doi.org/10.1103/PhysRevLett.126.056402} {\bibfield  {journal}
  {\bibinfo  {journal} {Phys. Rev. Lett.}\ }\textbf {\bibinfo {volume} {126}}
  (\bibinfo {year} {2021})}\BibitemShut {NoStop}%
\bibitem [{\citenamefont {White}\ and\ \citenamefont
  {Affleck}(2008)}]{White2008}%
  \BibitemOpen
  \bibfield  {author} {\bibinfo {author} {\bibfnamefont {S.~R.}\ \bibnamefont
  {White}}\ and\ \bibinfo {author} {\bibfnamefont {I.}~\bibnamefont
  {Affleck}},\ }\href {\doibase 10.1103/PhysRevB.77.134437} {\bibfield
  {journal} {\bibinfo  {journal} {Phys. Rev. B}\ }\textbf {\bibinfo {volume}
  {77}},\ \bibinfo {pages} {134437} (\bibinfo {year} {2008})}\BibitemShut
  {NoStop}%
\bibitem [{\citenamefont {Holzner}\ \emph {et~al.}(2011)\citenamefont
  {Holzner}, \citenamefont {Weichselbaum}, \citenamefont {McCulloch},
  \citenamefont {Schollw\"ock},\ and\ \citenamefont {von Delft}}]{Holzner2011}%
  \BibitemOpen
  \bibfield  {author} {\bibinfo {author} {\bibfnamefont {A.}~\bibnamefont
  {Holzner}}, \bibinfo {author} {\bibfnamefont {A.}~\bibnamefont
  {Weichselbaum}}, \bibinfo {author} {\bibfnamefont {I.~P.}\ \bibnamefont
  {McCulloch}}, \bibinfo {author} {\bibfnamefont {U.}~\bibnamefont
  {Schollw\"ock}}, \ and\ \bibinfo {author} {\bibfnamefont {J.}~\bibnamefont
  {von Delft}},\ }\href {\doibase 10.1103/PhysRevB.83.195115} {\bibfield
  {journal} {\bibinfo  {journal} {Phys. Rev. B}\ }\textbf {\bibinfo {volume}
  {83}},\ \bibinfo {pages} {195115} (\bibinfo {year} {2011})}\BibitemShut
  {NoStop}%
\bibitem [{\citenamefont {Nocera}\ and\ \citenamefont
  {Alvarez}(2016)}]{Nocera2016}%
  \BibitemOpen
  \bibfield  {author} {\bibinfo {author} {\bibfnamefont {A.}~\bibnamefont
  {Nocera}}\ and\ \bibinfo {author} {\bibfnamefont {G.}~\bibnamefont
  {Alvarez}},\ }\href {\doibase 10.1103/PhysRevE.94.053308} {\bibfield
  {journal} {\bibinfo  {journal} {Phys. Rev. E}\ }\textbf {\bibinfo {volume}
  {94}},\ \bibinfo {pages} {053308} (\bibinfo {year} {2016})}\BibitemShut
  {NoStop}%
\bibitem [{\citenamefont {Cheneau}\ \emph {et~al.}(2012)\citenamefont
  {Cheneau}, \citenamefont {Barmettler}, \citenamefont {Poletti}, \citenamefont
  {Endres}, \citenamefont {Schauß}, \citenamefont {Fukuhara}, \citenamefont
  {Gross}, \citenamefont {Bloch}, \citenamefont {Kollath},\ and\ \citenamefont
  {Kuhr}}]{Cheneau2012}%
  \BibitemOpen
  \bibfield  {author} {\bibinfo {author} {\bibfnamefont {M.}~\bibnamefont
  {Cheneau}}, \bibinfo {author} {\bibfnamefont {P.}~\bibnamefont {Barmettler}},
  \bibinfo {author} {\bibfnamefont {D.}~\bibnamefont {Poletti}}, \bibinfo
  {author} {\bibfnamefont {M.}~\bibnamefont {Endres}}, \bibinfo {author}
  {\bibfnamefont {P.}~\bibnamefont {Schauß}}, \bibinfo {author} {\bibfnamefont
  {T.}~\bibnamefont {Fukuhara}}, \bibinfo {author} {\bibfnamefont
  {C.}~\bibnamefont {Gross}}, \bibinfo {author} {\bibfnamefont
  {I.}~\bibnamefont {Bloch}}, \bibinfo {author} {\bibfnamefont
  {C.}~\bibnamefont {Kollath}}, \ and\ \bibinfo {author} {\bibfnamefont
  {S.}~\bibnamefont {Kuhr}},\ }\href {\doibase 10.1038/nature10748} {\bibfield
  {journal} {\bibinfo  {journal} {Nature}\ }\textbf {\bibinfo {volume} {481}},\
  \bibinfo {pages} {484} (\bibinfo {year} {2012})}\BibitemShut {NoStop}%
\bibitem [{\citenamefont {Meinert}\ \emph {et~al.}(2013)\citenamefont
  {Meinert}, \citenamefont {Mark}, \citenamefont {Kirilov}, \citenamefont
  {Lauber}, \citenamefont {Weinmann}, \citenamefont {Daley},\ and\
  \citenamefont {N\"agerl}}]{Meinert2013}%
  \BibitemOpen
  \bibfield  {author} {\bibinfo {author} {\bibfnamefont {F.}~\bibnamefont
  {Meinert}}, \bibinfo {author} {\bibfnamefont {M.~J.}\ \bibnamefont {Mark}},
  \bibinfo {author} {\bibfnamefont {E.}~\bibnamefont {Kirilov}}, \bibinfo
  {author} {\bibfnamefont {K.}~\bibnamefont {Lauber}}, \bibinfo {author}
  {\bibfnamefont {P.}~\bibnamefont {Weinmann}}, \bibinfo {author}
  {\bibfnamefont {A.~J.}\ \bibnamefont {Daley}}, \ and\ \bibinfo {author}
  {\bibfnamefont {H.-C.}\ \bibnamefont {N\"agerl}},\ }\href {\doibase
  10.1103/PhysRevLett.111.053003} {\bibfield  {journal} {\bibinfo  {journal}
  {Phys. Rev. Lett.}\ }\textbf {\bibinfo {volume} {111}},\ \bibinfo {pages}
  {053003} (\bibinfo {year} {2013})}\BibitemShut {NoStop}%
\bibitem [{\citenamefont {Langen}\ \emph {et~al.}(2013)\citenamefont {Langen},
  \citenamefont {Geiger}, \citenamefont {Kuhnert}, \citenamefont {Rauer},\ and\
  \citenamefont {Schmiedmayer}}]{Langen2013}%
  \BibitemOpen
  \bibfield  {author} {\bibinfo {author} {\bibfnamefont {T.}~\bibnamefont
  {Langen}}, \bibinfo {author} {\bibfnamefont {R.}~\bibnamefont {Geiger}},
  \bibinfo {author} {\bibfnamefont {M.}~\bibnamefont {Kuhnert}}, \bibinfo
  {author} {\bibfnamefont {B.}~\bibnamefont {Rauer}}, \ and\ \bibinfo {author}
  {\bibfnamefont {J.}~\bibnamefont {Schmiedmayer}},\ }\href {\doibase
  10.1038/nphys2739} {\bibfield  {journal} {\bibinfo  {journal} {Nat. Phys.}\
  }\textbf {\bibinfo {volume} {9}},\ \bibinfo {pages} {640} (\bibinfo {year}
  {2013})}\BibitemShut {NoStop}%
\bibitem [{\citenamefont {Kaufman}\ \emph {et~al.}(2016)\citenamefont
  {Kaufman}, \citenamefont {Tai}, \citenamefont {Lukin}, \citenamefont
  {Rispoli}, \citenamefont {Schittko}, \citenamefont {Preiss},\ and\
  \citenamefont {Greiner}}]{Kaufman2016}%
  \BibitemOpen
  \bibfield  {author} {\bibinfo {author} {\bibfnamefont {A.~M.}\ \bibnamefont
  {Kaufman}}, \bibinfo {author} {\bibfnamefont {M.~E.}\ \bibnamefont {Tai}},
  \bibinfo {author} {\bibfnamefont {A.}~\bibnamefont {Lukin}}, \bibinfo
  {author} {\bibfnamefont {M.}~\bibnamefont {Rispoli}}, \bibinfo {author}
  {\bibfnamefont {R.}~\bibnamefont {Schittko}}, \bibinfo {author}
  {\bibfnamefont {P.~M.}\ \bibnamefont {Preiss}}, \ and\ \bibinfo {author}
  {\bibfnamefont {M.}~\bibnamefont {Greiner}},\ }\href {\doibase
  10.1126/science.aaf6725} {\bibfield  {journal} {\bibinfo  {journal}
  {Science}\ }\textbf {\bibinfo {volume} {353}},\ \bibinfo {pages} {794}
  (\bibinfo {year} {2016})}\BibitemShut {NoStop}%
\bibitem [{\citenamefont {Calabrese}\ and\ \citenamefont
  {Cardy}(2005)}]{Calabrese2005}%
  \BibitemOpen
  \bibfield  {author} {\bibinfo {author} {\bibfnamefont {P.}~\bibnamefont
  {Calabrese}}\ and\ \bibinfo {author} {\bibfnamefont {J.}~\bibnamefont
  {Cardy}},\ }\href {\doibase 10.1088/1742-5468/2005/04/p04010} {\bibfield
  {journal} {\bibinfo  {journal} {J. Stat. Mech.: Theory Exp.}\ }\textbf
  {\bibinfo {volume} {2005}},\ \bibinfo {pages} {P04010} (\bibinfo {year}
  {2005})}\BibitemShut {NoStop}%
\bibitem [{\citenamefont {Ba\~nuls}\ \emph {et~al.}(2009)\citenamefont
  {Ba\~nuls}, \citenamefont {Hastings}, \citenamefont {Verstraete},\ and\
  \citenamefont {Cirac}}]{Banuls2009}%
  \BibitemOpen
  \bibfield  {author} {\bibinfo {author} {\bibfnamefont {M.~C.}\ \bibnamefont
  {Ba\~nuls}}, \bibinfo {author} {\bibfnamefont {M.~B.}\ \bibnamefont
  {Hastings}}, \bibinfo {author} {\bibfnamefont {F.}~\bibnamefont
  {Verstraete}}, \ and\ \bibinfo {author} {\bibfnamefont {J.~I.}\ \bibnamefont
  {Cirac}},\ }\href {\doibase 10.1103/PhysRevLett.102.240603} {\bibfield
  {journal} {\bibinfo  {journal} {Phys. Rev. Lett.}\ }\textbf {\bibinfo
  {volume} {102}},\ \bibinfo {pages} {240603} (\bibinfo {year}
  {2009})}\BibitemShut {NoStop}%
\bibitem [{\citenamefont {Müller-Hermes}\ \emph {et~al.}(2012)\citenamefont
  {Müller-Hermes}, \citenamefont {Cirac},\ and\ \citenamefont
  {Bañuls}}]{Mueller-Hermes2012}%
  \BibitemOpen
  \bibfield  {author} {\bibinfo {author} {\bibfnamefont {A.}~\bibnamefont
  {Müller-Hermes}}, \bibinfo {author} {\bibfnamefont {J.~I.}\ \bibnamefont
  {Cirac}}, \ and\ \bibinfo {author} {\bibfnamefont {M.~C.}\ \bibnamefont
  {Bañuls}},\ }\href {\doibase 10.1088/1367-2630/14/7/075003} {\bibfield
  {journal} {\bibinfo  {journal} {New J. Phys.}\ }\textbf {\bibinfo {volume}
  {14}},\ \bibinfo {pages} {075003} (\bibinfo {year} {2012})}\BibitemShut
  {NoStop}%
\bibitem [{\citenamefont {Fr\'{\i}as-P\'erez}\ and\ \citenamefont
  {Ba\~nuls}(2022)}]{Frias-Perez2022}%
  \BibitemOpen
  \bibfield  {author} {\bibinfo {author} {\bibfnamefont {M.}~\bibnamefont
  {Fr\'{\i}as-P\'erez}}\ and\ \bibinfo {author} {\bibfnamefont {M.~C.}\
  \bibnamefont {Ba\~nuls}},\ }\href {\doibase 10.1103/PhysRevB.106.115117}
  {\bibfield  {journal} {\bibinfo  {journal} {Phys. Rev. B}\ }\textbf {\bibinfo
  {volume} {106}},\ \bibinfo {pages} {115117} (\bibinfo {year}
  {2022})}\BibitemShut {NoStop}%
\bibitem [{\citenamefont {Lamb}\ \emph {et~al.}(2024)\citenamefont {Lamb},
  \citenamefont {Tang}, \citenamefont {Davis},\ and\ \citenamefont
  {Roy}}]{Lamb2024}%
  \BibitemOpen
  \bibfield  {author} {\bibinfo {author} {\bibfnamefont {C.}~\bibnamefont
  {Lamb}}, \bibinfo {author} {\bibfnamefont {Y.}~\bibnamefont {Tang}}, \bibinfo
  {author} {\bibfnamefont {R.}~\bibnamefont {Davis}}, \ and\ \bibinfo {author}
  {\bibfnamefont {A.}~\bibnamefont {Roy}},\ }\href {\doibase
  10.1038/s41467-024-50206-2} {\bibfield  {journal} {\bibinfo  {journal} {Nat.
  Commun.}\ }\textbf {\bibinfo {volume} {15}},\ \bibinfo {pages} {5901}
  (\bibinfo {year} {2024})}\BibitemShut {NoStop}%
\bibitem [{Note1()}]{Note1}%
  \BibitemOpen
  \bibinfo {note} {The pure fermionic Gaussian states are also known as
  Hartree-Fock-Bogoliubov states. In the absence of pairing, they reduce to
  Hartree-Fock states.}\BibitemShut {Stop}%
\bibitem [{\citenamefont {Zamolodchikov}(1989)}]{Zamolodchikov1989}%
  \BibitemOpen
  \bibfield  {author} {\bibinfo {author} {\bibfnamefont {A.~B.}\ \bibnamefont
  {Zamolodchikov}},\ }\href {\doibase 10.1142/S0217751X8900176X} {\bibfield
  {journal} {\bibinfo  {journal} {Int. J. Mod. Phys. A}\ }\textbf {\bibinfo
  {volume} {4}},\ \bibinfo {pages} {4235} (\bibinfo {year} {1989})}\BibitemShut
  {NoStop}%
\bibitem [{\citenamefont {Coldea}\ \emph {et~al.}(2010)\citenamefont {Coldea},
  \citenamefont {Tennant}, \citenamefont {Wheeler}, \citenamefont {Wawrzynska},
  \citenamefont {Prabhakaran}, \citenamefont {Telling}, \citenamefont
  {Habicht}, \citenamefont {Smeibidl},\ and\ \citenamefont
  {Kiefer}}]{Coldea2010}%
  \BibitemOpen
  \bibfield  {author} {\bibinfo {author} {\bibfnamefont {R.}~\bibnamefont
  {Coldea}}, \bibinfo {author} {\bibfnamefont {D.~A.}\ \bibnamefont {Tennant}},
  \bibinfo {author} {\bibfnamefont {E.~M.}\ \bibnamefont {Wheeler}}, \bibinfo
  {author} {\bibfnamefont {E.}~\bibnamefont {Wawrzynska}}, \bibinfo {author}
  {\bibfnamefont {D.}~\bibnamefont {Prabhakaran}}, \bibinfo {author}
  {\bibfnamefont {M.}~\bibnamefont {Telling}}, \bibinfo {author} {\bibfnamefont
  {K.}~\bibnamefont {Habicht}}, \bibinfo {author} {\bibfnamefont
  {P.}~\bibnamefont {Smeibidl}}, \ and\ \bibinfo {author} {\bibfnamefont
  {K.}~\bibnamefont {Kiefer}},\ }\href {\doibase 10.1126/science.1180085}
  {\bibfield  {journal} {\bibinfo  {journal} {Science}\ }\textbf {\bibinfo
  {volume} {327}},\ \bibinfo {pages} {177} (\bibinfo {year}
  {2010})}\BibitemShut {NoStop}%
\bibitem [{\citenamefont {Amelin}\ \emph {et~al.}(2020)\citenamefont {Amelin},
  \citenamefont {Engelmayer}, \citenamefont {Viirok}, \citenamefont {Nagel},
  \citenamefont {R\~o\ om}, \citenamefont {Lorenz},\ and\ \citenamefont
  {Wang}}]{Amelin2020}%
  \BibitemOpen
  \bibfield  {author} {\bibinfo {author} {\bibfnamefont {K.}~\bibnamefont
  {Amelin}}, \bibinfo {author} {\bibfnamefont {J.}~\bibnamefont {Engelmayer}},
  \bibinfo {author} {\bibfnamefont {J.}~\bibnamefont {Viirok}}, \bibinfo
  {author} {\bibfnamefont {U.}~\bibnamefont {Nagel}}, \bibinfo {author}
  {\bibfnamefont {T.}~\bibnamefont {R\~o\ om}}, \bibinfo {author}
  {\bibfnamefont {T.}~\bibnamefont {Lorenz}}, \ and\ \bibinfo {author}
  {\bibfnamefont {Z.}~\bibnamefont {Wang}},\ }\href {\doibase
  10.1103/PhysRevB.102.104431} {\bibfield  {journal} {\bibinfo  {journal}
  {Phys. Rev. B}\ }\textbf {\bibinfo {volume} {102}},\ \bibinfo {pages}
  {104431} (\bibinfo {year} {2020})}\BibitemShut {NoStop}%
\bibitem [{\citenamefont {Zhang}\ \emph {et~al.}(2020)\citenamefont {Zhang},
  \citenamefont {Amelin}, \citenamefont {Wang}, \citenamefont {Zou},
  \citenamefont {Yang}, \citenamefont {Nagel}, \citenamefont {R\~{o}\~{o}m},
  \citenamefont {Dey}, \citenamefont {Nugroho}, \citenamefont {Lorenz},
  \citenamefont {Wu},\ and\ \citenamefont {Wang}}]{ZhangZ2020}%
  \BibitemOpen
  \bibfield  {author} {\bibinfo {author} {\bibfnamefont {Z.}~\bibnamefont
  {Zhang}}, \bibinfo {author} {\bibfnamefont {K.}~\bibnamefont {Amelin}},
  \bibinfo {author} {\bibfnamefont {X.}~\bibnamefont {Wang}}, \bibinfo {author}
  {\bibfnamefont {H.}~\bibnamefont {Zou}}, \bibinfo {author} {\bibfnamefont
  {J.}~\bibnamefont {Yang}}, \bibinfo {author} {\bibfnamefont {U.}~\bibnamefont
  {Nagel}}, \bibinfo {author} {\bibfnamefont {T.}~\bibnamefont {R\~{o}\~{o}m}},
  \bibinfo {author} {\bibfnamefont {T.}~\bibnamefont {Dey}}, \bibinfo {author}
  {\bibfnamefont {A.~A.}\ \bibnamefont {Nugroho}}, \bibinfo {author}
  {\bibfnamefont {T.}~\bibnamefont {Lorenz}}, \bibinfo {author} {\bibfnamefont
  {J.}~\bibnamefont {Wu}}, \ and\ \bibinfo {author} {\bibfnamefont
  {Z.}~\bibnamefont {Wang}},\ }\href {\doibase 10.1103/PhysRevB.101.220411}
  {\bibfield  {journal} {\bibinfo  {journal} {Phys. Rev. B}\ }\textbf {\bibinfo
  {volume} {101}},\ \bibinfo {pages} {220411} (\bibinfo {year}
  {2020})}\BibitemShut {NoStop}%
\bibitem [{\citenamefont {Zou}\ \emph {et~al.}(2021)\citenamefont {Zou},
  \citenamefont {Cui}, \citenamefont {Wang}, \citenamefont {Zhang},
  \citenamefont {Yang}, \citenamefont {Xu}, \citenamefont {Okutani},
  \citenamefont {Hagiwara}, \citenamefont {Matsuda}, \citenamefont {Wang},
  \citenamefont {Mussardo}, \citenamefont {H\'ods\'agi}, \citenamefont
  {Kormos}, \citenamefont {He}, \citenamefont {Kimura}, \citenamefont {Yu},
  \citenamefont {Yu}, \citenamefont {Ma},\ and\ \citenamefont
  {Wu}}]{ZouHY2021}%
  \BibitemOpen
  \bibfield  {author} {\bibinfo {author} {\bibfnamefont {H.}~\bibnamefont
  {Zou}}, \bibinfo {author} {\bibfnamefont {Y.}~\bibnamefont {Cui}}, \bibinfo
  {author} {\bibfnamefont {X.}~\bibnamefont {Wang}}, \bibinfo {author}
  {\bibfnamefont {Z.}~\bibnamefont {Zhang}}, \bibinfo {author} {\bibfnamefont
  {J.}~\bibnamefont {Yang}}, \bibinfo {author} {\bibfnamefont {G.}~\bibnamefont
  {Xu}}, \bibinfo {author} {\bibfnamefont {A.}~\bibnamefont {Okutani}},
  \bibinfo {author} {\bibfnamefont {M.}~\bibnamefont {Hagiwara}}, \bibinfo
  {author} {\bibfnamefont {M.}~\bibnamefont {Matsuda}}, \bibinfo {author}
  {\bibfnamefont {G.}~\bibnamefont {Wang}}, \bibinfo {author} {\bibfnamefont
  {G.}~\bibnamefont {Mussardo}}, \bibinfo {author} {\bibfnamefont
  {K.}~\bibnamefont {H\'ods\'agi}}, \bibinfo {author} {\bibfnamefont
  {M.}~\bibnamefont {Kormos}}, \bibinfo {author} {\bibfnamefont
  {Z.}~\bibnamefont {He}}, \bibinfo {author} {\bibfnamefont {S.}~\bibnamefont
  {Kimura}}, \bibinfo {author} {\bibfnamefont {R.}~\bibnamefont {Yu}}, \bibinfo
  {author} {\bibfnamefont {W.}~\bibnamefont {Yu}}, \bibinfo {author}
  {\bibfnamefont {J.}~\bibnamefont {Ma}}, \ and\ \bibinfo {author}
  {\bibfnamefont {J.}~\bibnamefont {Wu}},\ }\href {\doibase
  10.1103/PhysRevLett.127.077201} {\bibfield  {journal} {\bibinfo  {journal}
  {Phys. Rev. Lett.}\ }\textbf {\bibinfo {volume} {127}},\ \bibinfo {pages}
  {077201} (\bibinfo {year} {2021})}\BibitemShut {NoStop}%
\bibitem [{\citenamefont {Amelin}\ \emph {et~al.}(2022)\citenamefont {Amelin},
  \citenamefont {Viirok}, \citenamefont {Nagel}, \citenamefont {Room},
  \citenamefont {Engelmayer}, \citenamefont {Dey}, \citenamefont {Nugroho},
  \citenamefont {Lorenz},\ and\ \citenamefont {Wang}}]{Amelin2022}%
  \BibitemOpen
  \bibfield  {author} {\bibinfo {author} {\bibfnamefont {K.}~\bibnamefont
  {Amelin}}, \bibinfo {author} {\bibfnamefont {J.}~\bibnamefont {Viirok}},
  \bibinfo {author} {\bibfnamefont {U.}~\bibnamefont {Nagel}}, \bibinfo
  {author} {\bibfnamefont {T.}~\bibnamefont {Room}}, \bibinfo {author}
  {\bibfnamefont {J.}~\bibnamefont {Engelmayer}}, \bibinfo {author}
  {\bibfnamefont {T.}~\bibnamefont {Dey}}, \bibinfo {author} {\bibfnamefont
  {A.~A.}\ \bibnamefont {Nugroho}}, \bibinfo {author} {\bibfnamefont
  {T.}~\bibnamefont {Lorenz}}, \ and\ \bibinfo {author} {\bibfnamefont
  {Z.}~\bibnamefont {Wang}},\ }\href {\doibase 10.1088/1751-8121/aca6b8}
  {\bibfield  {journal} {\bibinfo  {journal} {J. Phys. A: Math. Theor.}\
  }\textbf {\bibinfo {volume} {55}},\ \bibinfo {pages} {484005} (\bibinfo
  {year} {2022})}\BibitemShut {NoStop}%
\bibitem [{Note2()}]{Note2}%
  \BibitemOpen
  \bibinfo {note} {While $H_0$ is often part of $H$, one may also take $H_0$ as
  a ``mean-field'' approximation of $H$ and $V = H - H_0$.}\BibitemShut {Stop}%
\bibitem [{\citenamefont {Bravyi}(2005)}]{Bravyi2005}%
  \BibitemOpen
  \bibfield  {author} {\bibinfo {author} {\bibfnamefont {S.}~\bibnamefont
  {Bravyi}},\ }\href {\doibase 10.26421/QIC5.3-3} {\bibfield  {journal}
  {\bibinfo  {journal} {Quantum Inf. and Comp.}\ }\textbf {\bibinfo {volume}
  {5}},\ \bibinfo {pages} {216} (\bibinfo {year} {2005})}\BibitemShut {NoStop}%
\bibitem [{\citenamefont {Wilson}(1975)}]{Wilson1975}%
  \BibitemOpen
  \bibfield  {author} {\bibinfo {author} {\bibfnamefont {K.~G.}\ \bibnamefont
  {Wilson}},\ }\href
  {https://journals.aps.org/rmp/abstract/10.1103/RevModPhys.47.773} {\bibfield
  {journal} {\bibinfo  {journal} {Rev. Mod. Phys.}\ }\textbf {\bibinfo {volume}
  {47}},\ \bibinfo {pages} {773} (\bibinfo {year} {1975})}\BibitemShut
  {NoStop}%
\bibitem [{\citenamefont {Yurov}\ and\ \citenamefont
  {Zamolodchikov}(1990)}]{Yurov1990}%
  \BibitemOpen
  \bibfield  {author} {\bibinfo {author} {\bibfnamefont {V.~P.}\ \bibnamefont
  {Yurov}}\ and\ \bibinfo {author} {\bibfnamefont {A.~B.}\ \bibnamefont
  {Zamolodchikov}},\ }\href {\doibase 10.1142/S0217751X9000218X} {\bibfield
  {journal} {\bibinfo  {journal} {Int. J. Mod. Phys. A}\ }\textbf {\bibinfo
  {volume} {5}},\ \bibinfo {pages} {3221} (\bibinfo {year} {1990})}\BibitemShut
  {NoStop}%
\bibitem [{\citenamefont {Yurov}\ and\ \citenamefont
  {Zamolodchikov}(1991)}]{Yurov1991}%
  \BibitemOpen
  \bibfield  {author} {\bibinfo {author} {\bibfnamefont {V.~P.}\ \bibnamefont
  {Yurov}}\ and\ \bibinfo {author} {\bibfnamefont {A.~B.}\ \bibnamefont
  {Zamolodchikov}},\ }\href {\doibase 10.1142/S0217751X91002161} {\bibfield
  {journal} {\bibinfo  {journal} {Int. J. Mod. Phys. A}\ }\textbf {\bibinfo
  {volume} {6}},\ \bibinfo {pages} {4557} (\bibinfo {year} {1991})}\BibitemShut
  {NoStop}%
\bibitem [{\citenamefont {Lassig}\ \emph {et~al.}(1991)\citenamefont {Lassig},
  \citenamefont {Mussardo},\ and\ \citenamefont {Cardy}}]{Lassig1991}%
  \BibitemOpen
  \bibfield  {author} {\bibinfo {author} {\bibfnamefont {M.}~\bibnamefont
  {Lassig}}, \bibinfo {author} {\bibfnamefont {G.}~\bibnamefont {Mussardo}}, \
  and\ \bibinfo {author} {\bibfnamefont {J.~L.}\ \bibnamefont {Cardy}},\ }\href
  {\doibase 10.1016/0550-3213(91)90206-D} {\bibfield  {journal} {\bibinfo
  {journal} {Nucl. Phys. B}\ }\textbf {\bibinfo {volume} {348}},\ \bibinfo
  {pages} {591} (\bibinfo {year} {1991})}\BibitemShut {NoStop}%
\bibitem [{\citenamefont {Beria}\ \emph {et~al.}(2013)\citenamefont {Beria},
  \citenamefont {Brandino}, \citenamefont {Lepori}, \citenamefont {Konik},\
  and\ \citenamefont {Sierra}}]{Beria2013}%
  \BibitemOpen
  \bibfield  {author} {\bibinfo {author} {\bibfnamefont {M.}~\bibnamefont
  {Beria}}, \bibinfo {author} {\bibfnamefont {G.}~\bibnamefont {Brandino}},
  \bibinfo {author} {\bibfnamefont {L.}~\bibnamefont {Lepori}}, \bibinfo
  {author} {\bibfnamefont {R.}~\bibnamefont {Konik}}, \ and\ \bibinfo {author}
  {\bibfnamefont {G.}~\bibnamefont {Sierra}},\ }\href {\doibase
  https://doi.org/10.1016/j.nuclphysb.2013.10.005} {\bibfield  {journal}
  {\bibinfo  {journal} {Nucl. Phys. B}\ }\textbf {\bibinfo {volume} {877}},\
  \bibinfo {pages} {457} (\bibinfo {year} {2013})}\BibitemShut {NoStop}%
\bibitem [{\citenamefont {Konik}\ \emph {et~al.}(2015)\citenamefont {Konik},
  \citenamefont {Pálmai}, \citenamefont {Takács},\ and\ \citenamefont
  {Tsvelik}}]{Konik2015}%
  \BibitemOpen
  \bibfield  {author} {\bibinfo {author} {\bibfnamefont {R.}~\bibnamefont
  {Konik}}, \bibinfo {author} {\bibfnamefont {T.}~\bibnamefont {Pálmai}},
  \bibinfo {author} {\bibfnamefont {G.}~\bibnamefont {Takács}}, \ and\
  \bibinfo {author} {\bibfnamefont {A.}~\bibnamefont {Tsvelik}},\ }\href
  {\doibase https://doi.org/10.1016/j.nuclphysb.2015.08.016} {\bibfield
  {journal} {\bibinfo  {journal} {Nucl. Phys. B}\ }\textbf {\bibinfo {volume}
  {899}},\ \bibinfo {pages} {547} (\bibinfo {year} {2015})}\BibitemShut
  {NoStop}%
\bibitem [{\citenamefont {James}\ \emph {et~al.}(2018)\citenamefont {James},
  \citenamefont {Konik}, \citenamefont {Lecheminant}, \citenamefont
  {Robinson},\ and\ \citenamefont {Tsvelik}}]{James2018}%
  \BibitemOpen
  \bibfield  {author} {\bibinfo {author} {\bibfnamefont {A.~J.~A.}\
  \bibnamefont {James}}, \bibinfo {author} {\bibfnamefont {R.~M.}\ \bibnamefont
  {Konik}}, \bibinfo {author} {\bibfnamefont {P.}~\bibnamefont {Lecheminant}},
  \bibinfo {author} {\bibfnamefont {N.~J.}\ \bibnamefont {Robinson}}, \ and\
  \bibinfo {author} {\bibfnamefont {A.~M.}\ \bibnamefont {Tsvelik}},\ }\href
  {\doibase 10.1088/1361-6633/aa91ea} {\bibfield  {journal} {\bibinfo
  {journal} {Rep. Prog. Phys.}\ }\textbf {\bibinfo {volume} {81}},\ \bibinfo
  {pages} {046002} (\bibinfo {year} {2018})}\BibitemShut {NoStop}%
\bibitem [{\citenamefont {Mussardo}(2020)}]{Mussardo-Book}%
  \BibitemOpen
  \bibfield  {author} {\bibinfo {author} {\bibfnamefont {G.}~\bibnamefont
  {Mussardo}},\ }\href {\doibase 10.1093/oso/9780198788102.001.0001} {\emph
  {\bibinfo {title} {Statistical Field Theory: An Introduction to Exactly
  Solved Models in Statistical Physics}}}\ (\bibinfo  {publisher} {Oxford
  University Press, Oxford},\ \bibinfo {year} {2020})\BibitemShut {NoStop}%
\bibitem [{\citenamefont {Pfeuty}(1970)}]{Pfeuty1970}%
  \BibitemOpen
  \bibfield  {author} {\bibinfo {author} {\bibfnamefont {P.}~\bibnamefont
  {Pfeuty}},\ }\href {\doibase https://doi.org/10.1016/0003-4916(70)90270-8}
  {\bibfield  {journal} {\bibinfo  {journal} {Ann. Phys.}\ }\textbf {\bibinfo
  {volume} {57}},\ \bibinfo {pages} {79} (\bibinfo {year} {1970})}\BibitemShut
  {NoStop}%
\bibitem [{Note3()}]{Note3}%
  \BibitemOpen
  \bibinfo {note} {For simplicity, we use the same truncation dimension $\chi $
  across all momentum sectors.}\BibitemShut {Stop}%
\bibitem [{App()}]{Append}%
  \BibitemOpen
  \href@noop {} {\bibinfo  {journal} {See the Appendices for further details
  about the Pfaffian formula for computing matrix elements in fermionic
  Gaussian basis, truncation in quasiparticle number in the simulation of
  quench dynamics, and the Fourier transform (power spectrum) of the
  time-evolved longitudinal magnetization, including
  Refs.~\cite{Bloch1962,Calabrese2011,Calabrese2012,Kormos2017}}\ }\BibitemShut
  {NoStop}%
\bibitem [{\citenamefont {Bertsch}\ and\ \citenamefont
  {Robledo}(2012)}]{Bertsch2012}%
  \BibitemOpen
\bibfield  {journal} {  }\bibfield  {author} {\bibinfo {author} {\bibfnamefont
  {G.~F.}\ \bibnamefont {Bertsch}}\ and\ \bibinfo {author} {\bibfnamefont
  {L.~M.}\ \bibnamefont {Robledo}},\ }\href {\doibase
  10.1103/PhysRevLett.108.042505} {\bibfield  {journal} {\bibinfo  {journal}
  {Phys. Rev. Lett.}\ }\textbf {\bibinfo {volume} {108}},\ \bibinfo {pages}
  {042505} (\bibinfo {year} {2012})}\BibitemShut {NoStop}%
\bibitem [{\citenamefont {Carlsson}\ and\ \citenamefont
  {Rotureau}(2021)}]{Carlsson2021}%
  \BibitemOpen
  \bibfield  {author} {\bibinfo {author} {\bibfnamefont {B.~G.}\ \bibnamefont
  {Carlsson}}\ and\ \bibinfo {author} {\bibfnamefont {J.}~\bibnamefont
  {Rotureau}},\ }\href {\doibase 10.1103/PhysRevLett.126.172501} {\bibfield
  {journal} {\bibinfo  {journal} {Phys. Rev. Lett.}\ }\textbf {\bibinfo
  {volume} {126}},\ \bibinfo {pages} {172501} (\bibinfo {year}
  {2021})}\BibitemShut {NoStop}%
\bibitem [{\citenamefont {Jin}\ \emph {et~al.}(2022)\citenamefont {Jin},
  \citenamefont {Sun}, \citenamefont {Zhou},\ and\ \citenamefont
  {Tu}}]{JinHK2022}%
  \BibitemOpen
  \bibfield  {author} {\bibinfo {author} {\bibfnamefont {H.-K.}\ \bibnamefont
  {Jin}}, \bibinfo {author} {\bibfnamefont {R.-Y.}\ \bibnamefont {Sun}},
  \bibinfo {author} {\bibfnamefont {Y.}~\bibnamefont {Zhou}}, \ and\ \bibinfo
  {author} {\bibfnamefont {H.-H.}\ \bibnamefont {Tu}},\ }\href {\doibase
  10.1103/PhysRevB.105.L081101} {\bibfield  {journal} {\bibinfo  {journal}
  {Phys. Rev. B}\ }\textbf {\bibinfo {volume} {105}},\ \bibinfo {pages}
  {L081101} (\bibinfo {year} {2022})}\BibitemShut {NoStop}%
\bibitem [{\citenamefont {Mascot}\ \emph {et~al.}(2023)\citenamefont {Mascot},
  \citenamefont {Hodge}, \citenamefont {Crawford}, \citenamefont {Bedow},
  \citenamefont {Morr},\ and\ \citenamefont {Rachel}}]{Mascot2023}%
  \BibitemOpen
  \bibfield  {author} {\bibinfo {author} {\bibfnamefont {E.}~\bibnamefont
  {Mascot}}, \bibinfo {author} {\bibfnamefont {T.}~\bibnamefont {Hodge}},
  \bibinfo {author} {\bibfnamefont {D.}~\bibnamefont {Crawford}}, \bibinfo
  {author} {\bibfnamefont {J.}~\bibnamefont {Bedow}}, \bibinfo {author}
  {\bibfnamefont {D.~K.}\ \bibnamefont {Morr}}, \ and\ \bibinfo {author}
  {\bibfnamefont {S.}~\bibnamefont {Rachel}},\ }\href {\doibase
  10.1103/PhysRevLett.131.176601} {\bibfield  {journal} {\bibinfo  {journal}
  {Phys. Rev. Lett.}\ }\textbf {\bibinfo {volume} {131}},\ \bibinfo {pages}
  {176601} (\bibinfo {year} {2023})}\BibitemShut {NoStop}%
\bibitem [{\citenamefont {Wimmer}(2012)}]{Wimmer2012}%
  \BibitemOpen
  \bibfield  {author} {\bibinfo {author} {\bibfnamefont {M.}~\bibnamefont
  {Wimmer}},\ }\href {\doibase 10.1145/2331130.2331138} {\bibfield  {journal}
  {\bibinfo  {journal} {ACM Trans. Math. Softw.}\ }\textbf {\bibinfo {volume}
  {38}},\ \bibinfo {pages} {1} (\bibinfo {year} {2012})}\BibitemShut {NoStop}%
\bibitem [{\citenamefont {Kj\"all}\ \emph {et~al.}(2011)\citenamefont
  {Kj\"all}, \citenamefont {Pollmann},\ and\ \citenamefont
  {Moore}}]{Kjaell2011}%
  \BibitemOpen
  \bibfield  {author} {\bibinfo {author} {\bibfnamefont {J.~A.}\ \bibnamefont
  {Kj\"all}}, \bibinfo {author} {\bibfnamefont {F.}~\bibnamefont {Pollmann}}, \
  and\ \bibinfo {author} {\bibfnamefont {J.~E.}\ \bibnamefont {Moore}},\ }\href
  {\doibase 10.1103/PhysRevB.83.020407} {\bibfield  {journal} {\bibinfo
  {journal} {Phys. Rev. B}\ }\textbf {\bibinfo {volume} {83}},\ \bibinfo
  {pages} {020407} (\bibinfo {year} {2011})}\BibitemShut {NoStop}%
\bibitem [{Note4()}]{Note4}%
  \BibitemOpen
  \bibinfo {note} {To ensure the results are not dominated by finite-size
  effects, the correlation length $\xi \sim m_1^{-1} \sim h_\parallel ^{-8/15}$
  should be small compared to the length of the chain. We find that for
  $h_\parallel = 0.05$ this is still the case.}\BibitemShut {Stop}%
\bibitem [{\citenamefont {Rakovszky}\ \emph {et~al.}(2016)\citenamefont
  {Rakovszky}, \citenamefont {Mestyán}, \citenamefont {Collura}, \citenamefont
  {Kormos},\ and\ \citenamefont {Takács}}]{Rakovszky2016}%
  \BibitemOpen
  \bibfield  {author} {\bibinfo {author} {\bibfnamefont {T.}~\bibnamefont
  {Rakovszky}}, \bibinfo {author} {\bibfnamefont {M.}~\bibnamefont {Mestyán}},
  \bibinfo {author} {\bibfnamefont {M.}~\bibnamefont {Collura}}, \bibinfo
  {author} {\bibfnamefont {M.}~\bibnamefont {Kormos}}, \ and\ \bibinfo {author}
  {\bibfnamefont {G.}~\bibnamefont {Takács}},\ }\href {\doibase
  https://doi.org/10.1016/j.nuclphysb.2016.08.024} {\bibfield  {journal}
  {\bibinfo  {journal} {Nucl. Phys. B}\ }\textbf {\bibinfo {volume} {911}},\
  \bibinfo {pages} {805} (\bibinfo {year} {2016})}\BibitemShut {NoStop}%
\bibitem [{\citenamefont {Kukuljan}\ \emph {et~al.}(2018)\citenamefont
  {Kukuljan}, \citenamefont {Sotiriadis},\ and\ \citenamefont
  {Takacs}}]{Kukuljan2018}%
  \BibitemOpen
  \bibfield  {author} {\bibinfo {author} {\bibfnamefont {I.}~\bibnamefont
  {Kukuljan}}, \bibinfo {author} {\bibfnamefont {S.}~\bibnamefont
  {Sotiriadis}}, \ and\ \bibinfo {author} {\bibfnamefont {G.}~\bibnamefont
  {Takacs}},\ }\href {\doibase 10.1103/PhysRevLett.121.110402} {\bibfield
  {journal} {\bibinfo  {journal} {Phys. Rev. Lett.}\ }\textbf {\bibinfo
  {volume} {121}},\ \bibinfo {pages} {110402} (\bibinfo {year}
  {2018})}\BibitemShut {NoStop}%
\bibitem [{\citenamefont {Hódsági}\ \emph {et~al.}(2018)\citenamefont
  {Hódsági}, \citenamefont {Kormos},\ and\ \citenamefont
  {Takács}}]{Hodsagi2018}%
  \BibitemOpen
  \bibfield  {author} {\bibinfo {author} {\bibfnamefont {K.}~\bibnamefont
  {Hódsági}}, \bibinfo {author} {\bibfnamefont {M.}~\bibnamefont {Kormos}}, \
  and\ \bibinfo {author} {\bibfnamefont {G.}~\bibnamefont {Takács}},\ }\href
  {\doibase 10.21468/SciPostPhys.5.3.027} {\bibfield  {journal} {\bibinfo
  {journal} {SciPost Phys.}\ }\textbf {\bibinfo {volume} {5}},\ \bibinfo
  {pages} {027} (\bibinfo {year} {2018})}\BibitemShut {NoStop}%
\bibitem [{\citenamefont {Konik}\ and\ \citenamefont
  {Adamov}(2007)}]{Konik2007}%
  \BibitemOpen
  \bibfield  {author} {\bibinfo {author} {\bibfnamefont {R.~M.}\ \bibnamefont
  {Konik}}\ and\ \bibinfo {author} {\bibfnamefont {Y.}~\bibnamefont {Adamov}},\
  }\href {\doibase 10.1103/PhysRevLett.98.147205} {\bibfield  {journal}
  {\bibinfo  {journal} {Phys. Rev. Lett.}\ }\textbf {\bibinfo {volume} {98}},\
  \bibinfo {pages} {147205} (\bibinfo {year} {2007})}\BibitemShut {NoStop}%
\bibitem [{Note5()}]{Note5}%
  \BibitemOpen
  \bibinfo {note} {\protect \url
  {www.nhr-verein.de/unsere-partner}}\BibitemShut {NoStop}%
\bibitem [{\citenamefont {Bloch}\ and\ \citenamefont
  {Messiah}(1962)}]{Bloch1962}%
  \BibitemOpen
  \bibfield  {author} {\bibinfo {author} {\bibfnamefont {C.}~\bibnamefont
  {Bloch}}\ and\ \bibinfo {author} {\bibfnamefont {A.}~\bibnamefont
  {Messiah}},\ }\href {\doibase https://doi.org/10.1016/0029-5582(62)90377-2}
  {\bibfield  {journal} {\bibinfo  {journal} {Nucl. Phys.}\ }\textbf {\bibinfo
  {volume} {39}},\ \bibinfo {pages} {95} (\bibinfo {year} {1962})}\BibitemShut
  {NoStop}%
\bibitem [{\citenamefont {Calabrese}\ \emph {et~al.}(2011)\citenamefont
  {Calabrese}, \citenamefont {Essler},\ and\ \citenamefont
  {Fagotti}}]{Calabrese2011}%
  \BibitemOpen
  \bibfield  {author} {\bibinfo {author} {\bibfnamefont {P.}~\bibnamefont
  {Calabrese}}, \bibinfo {author} {\bibfnamefont {F.~H.~L.}\ \bibnamefont
  {Essler}}, \ and\ \bibinfo {author} {\bibfnamefont {M.}~\bibnamefont
  {Fagotti}},\ }\href {\doibase 10.1103/PhysRevLett.106.227203} {\bibfield
  {journal} {\bibinfo  {journal} {Phys. Rev. Lett.}\ }\textbf {\bibinfo
  {volume} {106}},\ \bibinfo {pages} {227203} (\bibinfo {year}
  {2011})}\BibitemShut {NoStop}%
\bibitem [{\citenamefont {Calabrese}\ \emph {et~al.}(2012)\citenamefont
  {Calabrese}, \citenamefont {Essler},\ and\ \citenamefont
  {Fagotti}}]{Calabrese2012}%
  \BibitemOpen
  \bibfield  {author} {\bibinfo {author} {\bibfnamefont {P.}~\bibnamefont
  {Calabrese}}, \bibinfo {author} {\bibfnamefont {F.~H.~L.}\ \bibnamefont
  {Essler}}, \ and\ \bibinfo {author} {\bibfnamefont {M.}~\bibnamefont
  {Fagotti}},\ }\href {\doibase 10.1088/1742-5468/2012/07/p07016} {\bibfield
  {journal} {\bibinfo  {journal} {J. Stat. Mech.: Theory Exp.}\ }\textbf
  {\bibinfo {volume} {2012}},\ \bibinfo {pages} {P07016} (\bibinfo {year}
  {2012})}\BibitemShut {NoStop}%
\bibitem [{\citenamefont {Kormos}\ \emph {et~al.}(2017)\citenamefont {Kormos},
  \citenamefont {Collura}, \citenamefont {Takács},\ and\ \citenamefont
  {Calabrese}}]{Kormos2017}%
  \BibitemOpen
  \bibfield  {author} {\bibinfo {author} {\bibfnamefont {M.}~\bibnamefont
  {Kormos}}, \bibinfo {author} {\bibfnamefont {M.}~\bibnamefont {Collura}},
  \bibinfo {author} {\bibfnamefont {G.}~\bibnamefont {Takács}}, \ and\
  \bibinfo {author} {\bibfnamefont {P.}~\bibnamefont {Calabrese}},\ }\href
  {\doibase 10.1038/nphys3934} {\bibfield  {journal} {\bibinfo  {journal} {Nat.
  Phys.}\ }\textbf {\bibinfo {volume} {13}},\ \bibinfo {pages} {246} (\bibinfo
  {year} {2017})}\BibitemShut {NoStop}%
\bibitem [{Note6()}]{Note6}%
  \BibitemOpen
  \bibinfo {note} {If the translation symmetry were absent, one would have to
  compute the matrix elements of each $\sigma ^x_j$ separately using $\sigma
  ^x_j = (c_j+c_j^\dagger ) \DOTSB \prod@ \slimits@ _{l=1}^{j-1} [c_l^{\dagger
  } + c_l)(c_l^{\dagger } - c_l)]$.}\BibitemShut {Stop}%
\end{thebibliography}%
\bibliographystyle{apsrev4-1}

\clearpage
\onecolumngrid

\section*{Appendix}

\setcounter{table}{0}
\renewcommand{\thetable}{S\arabic{table}}
\setcounter{figure}{0}
\renewcommand{\thefigure}{S\arabic{figure}}
\setcounter{equation}{0}
\renewcommand{\theequation}{S\arabic{equation}}

\appendix

\section{Pfaffian formula for matrix elements in fermionic Gaussian basis}

In the Truncated Gaussian Basis Approach (TGBA), a key technical step is the evaluation of operator matrix elements in the basis of fermionic Gaussian states. This step is essential both for constructing the effective Hamiltonian within the truncated subspace and for computing physical observables in the variational solution, which is expressed as a superposition of fermionic Gaussian states. In the following, we detail the procedure for such computations.

We begin with a brief review of Wick's theorem, which is essential for performing the computation. Let us consider $N$ fermionic modes, denoted by creation (annihilation) operators $c_j^\dagger$ ($c_j$) with $j=1,\ldots,N$. The vacuum of these fermionic modes is written as $|0\rangle_c$, satisfying $c_j |0\rangle_c = 0 \ \forall j$. Now, consider $2n$ operators $\mathcal{O}_m$ ($m=1,\ldots,2n$), which are all expressed as linear combinations of the fermionic creation and annihilation operators, $\mathcal{O}_m = \sum_{j=1}^N (U_{jm} c_j^\dagger + V_{jm} c_j)$, where $U$ and $V$ are $N \times 2n$ matrices. Then, Wick's theorem states that the expectation value of the product of these operators in $|0\rangle_c$ can be written as the Pfaffian of a $2n \times 2n$ matrix $A$:
\begin{align}\label{eq:Pfaffian}
    _{c} \langle 0| \mathcal{O}_1 \cdots \mathcal{O}_{2n} |0\rangle_c = \mathrm{Pf}(A) \, ,
\end{align}
where $A$ is skew-symmetric, and its upper triangular part is given by the upper triangular (abbreviated as ``$\mathrm{u.t.}$'' hereafter) part $(V^T U)|_\mathrm{u.t.}$ of $V^T U$:
\begin{align}
    A =
    \begin{pmatrix}
        0 & & (V^T U)|_\mathrm{u.t.} \\
        & \ddots & \\
        - [(V^T U)|_\mathrm{u.t.}]^T & & 0
    \end{pmatrix}.
\label{eq:Pfaffian-SM}
\end{align}

For the transverse-field Ising chain (TFIC) considered in the main text [$H_0$ in Eq.~(3)], the eigenstates in both the NS- and the R-sector are written as an even number of Bogoliubov modes $d_k^\dagger = \sin(\theta_k/2) c_k^\dagger + i \cos(\theta_k/2) c_{-k}$ applied to Bogoliubov vacua $|0\rangle_\mathrm{NS/R}$ ($d_k |0\rangle_\mathrm{NS/R} = 0  \; \forall k \in \mathrm{NS/R}$), where $\theta_k  = \mathrm{sign}(k) \, \arccos \frac{2(h_\perp + \cos k)}{\varepsilon_k} \in (-\pi,\pi]$ is the Bogoliubov angle. Moreover, the Bogoliubov vacua $|0\rangle_\mathrm{NS/R}$ can be expressed in terms of the Bogoliubov modes applied to the vacuum of the Jordan-Wigner fermions:
\begin{align}
    |0\rangle_\mathrm{NS/R} = \prod_{0< k \in \mathrm{NS/R}} \left[\cos(\theta_k/2)\right]^{-1} \prod_{k \in \mathrm{NS/R}, k \neq \pi} d_k |0\rangle_c \, ,
\label{eq:TFIC-vacua-SM}
\end{align}
where $|0\rangle_c$ is now the vacuum of the Jordan-Wigner fermions ($c_j |0\rangle_c = 0 \; \forall j$). Since Bogoliubov modes are themselves linear combinations of fermonic creation and annihilation operators, an operator that can be written as a linear combination of fermionic creation and annihilation operators, such as $\mathcal{O}_m$ discussed above, can have its matrix elements evaluated in the TFIC eigenbasis using the Pfaffian formula given in Eq.~\eqref{eq:Pfaffian-SM}. Similarly, the matrix elements of products of such operators, can also be computed using the same method.

Since the number of Bogoliubov modes in $|0\rangle_\mathrm{NS/R}$ grows linearly with the system size, the size of the matrix $A$ also grows linearly ($2n \sim \mathcal{O}(N)$) when calculating overlaps/matrix elements with the TFIC eigenstates. Current algorithms can calculate Pfaffians in $\mathcal{O}((2n)^3)$ time~\cite{Wimmer2012}, enabling us to set up the truncated Hamiltonian efficiently for system sizes up to $N \sim 100$ sites.

For the quantum Ising chain with translation symmetry [Eq.~(3) in the main text], evaluating the matrix elements $V = - h_\parallel \sum_{j=1}^N \sigma_j^x$ in the TFIC eigenbasis can be simplified significantly~\footnote{If the translation symmetry were absent, one would have to compute the matrix elements of each $\sigma^x_j$ separately using $\sigma^x_j = (c_j+c_j^\dagger) \prod_{l=1}^{j-1} [c_l^{\dagger} + c_l)(c_l^{\dagger} - c_l)]$.}. Since $V$ is translation-invariant, it does not mix different momentum sectors of the truncated subspace, enabling efficient calculations through parallelization. Translation invariance further implies that only the matrix elements of $\sigma^x_{1}$ need to be evaluated, yielding the following formula:
\begin{align}
    \langle \phi_{k, \alpha}|V|\phi_{k, \beta}\rangle = -N h_\parallel \langle \phi_{k,\alpha}|(c^\dagger_1 + c_1)|\phi_{k, \beta} \rangle \, .
\label{eq:V-matrix-element-SM}
\end{align}
Note also that because $\sigma_1^x = c^\dagger_1 + c_1$ changes the fermion parity, the matrix elements in Eq.~\eqref{eq:V-matrix-element-SM} are non-vanishing only when $|\phi_{k,\alpha}\rangle$ and $ |\phi_{k,\beta}\rangle$ have opposite fermion parity (one lies in the NS sector and the other in the R sector).

As a final remark in this technical section, we note that the techniques described above are readily extended to other systems within the TGBA framework. For example, using the Bloch-Messiah decomposition~\cite{Bloch1962}, any fermionic Gaussian state can be expressed as a number of Bogoliubov operators acting on the original fermionic vacuum~\cite{JinHK2022}, thereby generalizing Eq.~\eqref{eq:TFIC-vacua-SM}.

\section{Truncation in quasiparticle number and quench dynamics}

In this section, we elaborate further on the truncation scheme in the simulation of quench dynamics. The fermionic Gaussian states in the full Gaussian basis are written as $d_{k_1}^\dagger \cdots d_{k_M}^\dagger |0\rangle_d$, where $|0\rangle_d$ is the Bogolubov vacuum, $d_{k}^\dagger$'s are Bogoliubov quasiparticle operators, and $M$ denotes the number of Bogoliubov quasiparticles.
When constructing the truncated subspace $\mathcal{H}_{\mathrm{trunc}}$, we first truncate in terms of the number of Bogoliubov quasiparticles, $M$. If the truncated subspace is constructed by sorting states according to the (bare) single-particle energy, as was done for the calculation of the dynamic structure factor in the main text, this additional truncation is natural since states with more quasiparticles tend to have higher energies. However, when simulating quench dynamics, where the overlap with the initial state $|\Phi_0\rangle$ is the more suitable metric, some additional care needs to be taken: Eigenstates of \emph{different} Hamiltonians can have non-negligible overlaps even when they differ drastically in (bare) energy and quasiparticle number.

In the TGBA framework, one can numerically calculate the overlaps of eigenstates of different Hamiltonians using the Pfaffian formula~\eqref{eq:Pfaffian}. Actually, if one considers the TFIC without the longitudinal field and only quenches the transverse field, these overlaps can be calculated analytically~\cite{Calabrese2011, Calabrese2012}, which provide useful analytical insights. In this case, the initial state, which is a superposition of the NS- and R-vacua $|0, h_\perp(t=0)\rangle_\text{NS,R}$ of the pre-quench Hamiltonian at $t=0$, can be written in terms of the vacua and quasiparticle operators of the post-quench Hamiltonian, with
\begin{align}
    |0, h_\perp(t=0)\rangle_{\text{NS,R}} &= \frac{1}{\mathcal{N}_{\text{NS,R}}} \prod_{0< k \in \text{NS,R} < \pi} \left[ 1 + i W(k) d_{-k}^\dagger d_k^\dagger \right] |0, h_\perp(t>0) \rangle_\text{NS,R} \, .
\label{eq:Calabrese_formula-SM}
\end{align}
Here $\mathcal{N}_{\text{NS,R}}$ are normalization factors, $d_k^\dagger$ are the Bogoliubov quasiparticle operators of the post-quench Hamiltonian, and the function $W(k) = \tan \frac{\Delta_k}{2} $ depends on the difference $\Delta_k$ of the Bogoliubov angles corresponding to the pre-quench and post-quench Hamiltonians, $\Delta_k = \theta_k(t>0) - \theta_k(t=0)$. Expanding the product in Eq.~\eqref{eq:Calabrese_formula-SM} in powers of $W(k)$ corresponds to an expansion in terms of the number of quasiparticles of the post-quench Hamiltonian:
\begin{align}
    |0, h_\perp(t=0) \rangle_\text{NS,R} &= \frac{1}{\mathcal{N}_\text{NS,R}} \left[ 1 + i \sum_{0<k \in \text{NS,R} < \pi} W(k) d_{-k}^\dagger d_k^\dagger - \sum_{0 < k,k' \in \text{NS,R} < \pi} W(k) W(k') d_{-k}^\dagger d_k^\dagger d_{-k'}^\dagger d_{k'}^\dagger + \mathcal{O}(W^3) \right]  \nonumber \\
    &\phantom{=} \;\times |0, h_\perp(t>0)\rangle_\text{NS,R} \, .
\end{align}
If the difference in the Bogoliubov angles $\Delta_k$ is small for all $k$, which is the case when one quenches not close to the critical point, higher-order terms in $W$ can be neglected and only states with a small number of quasiparticles need to be taken into account. For quenches well within the ferromagnetic or paramagnetic phase of the TFIC, we observed that retaining states with up to six quasiparticles is usually sufficient.

\begin{figure}[h]
    \resizebox{\columnwidth}{!}{\includegraphics{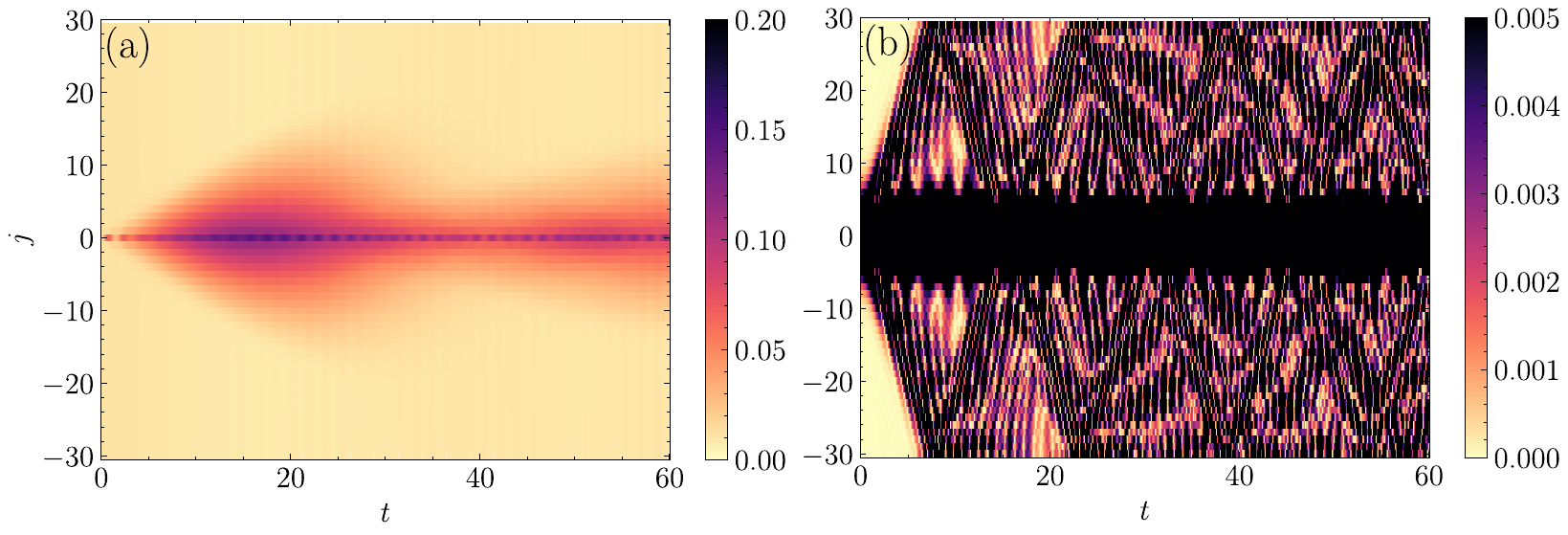}}
\caption{Dynamic correlation function $\langle \sigma_1^x(t) \sigma_{j+1}^x(t) \rangle - \langle \sigma_1^x(t) \rangle \langle \sigma_{j+1}^x(t) \rangle$ for two different quenches (a) $(h_\perp, h_\parallel) = (0, 0) \mapsto (0.25, 0.05)$ and (b) $(h_\perp, h_\parallel) = (2.0, 0) \mapsto (1.75, 0.05)$. For both simulations, the system size is $N=60$ and $\chi = 2000$ states were retained in the truncated subspace.}
\label{fig:cf}
\end{figure}

These considerations give a hint that if the quench is not too close to the critical point, the TGBA can simulate quench dynamics by performing a truncation with respect to the quasiparticle number in the fermionic Gaussian basis. This insight is quite useful in reducing the complexity in constructing $\mathcal{H}_{\mathrm{trunc}}$ and simulating computationally expensive quantities, such as dynamic correlation functions. For the quantum Ising chain with both transverse and longitudinal fields present, Figure~\ref{fig:cf} displays the results for the equal-time correlation function $\langle \sigma_1^x(t) \sigma_{j+1}^x(t) \rangle - \langle \sigma_1^x(t) \rangle \langle \sigma_{j+1}^x(t) \rangle$ for two example quenches, where for one of them, the initial state is within the ferromagnetic phase of the TFIC and the other lies within the paramagnetic phase. The results are in agreement with the expected behavior that excitations in the paramagnetic regime spread with a light-cone structure, while in the ferromagnetic regime the longitudinal field confines excitations after the quench, leading to correlations being mostly confined to a bounded region~\cite{Kormos2017}.

If the quench is close to the critical point, the truncation with respect to the quasiparticle number is not guaranteed to be optimal. However, for the quench from the ferromagnetic phase to the $E_8$ regime (see main text), we observe that this truncation scheme still provides very accurate results when moderate number of quasiparticles ($8 \sim 10$) are retained.

\section{Power spectrum and the $E_8$ masses}

In this section, we provide numerical data for the Fourier transform of the time evolution of longitudinal magnetization $\langle \sigma^x \rangle (t)  \equiv \frac{1}{N} \sum_{j=1}^N\langle \sigma^x_j \rangle(t)$ following the sudden quench (see description in the main text). The Fourier transform is defined as $\langle \sigma^x\rangle(\omega) = \int_{-\infty}^{+\infty} \mathrm{d}t \, e^{-i\omega t} \, \langle \sigma^x \rangle (t)$.

\begin{figure}[h]
    \includegraphics[width=0.5\textwidth]{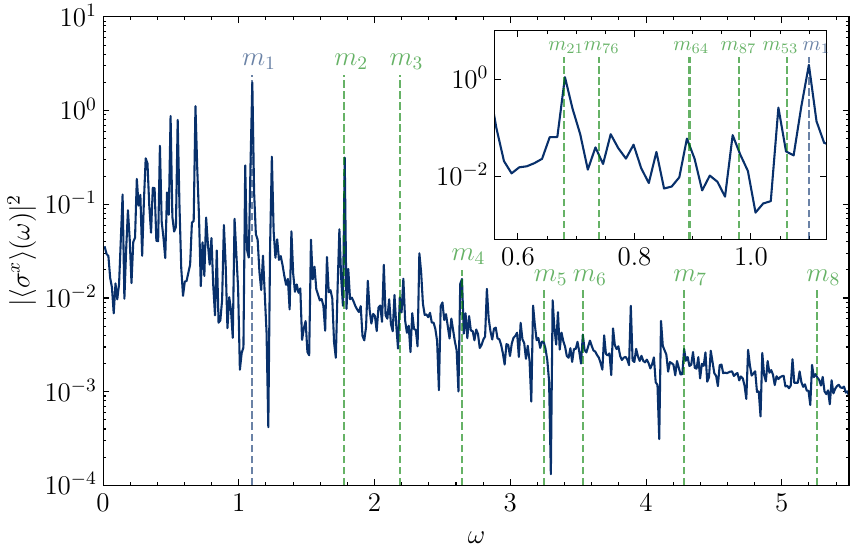}
    \caption{Power spectrum $|\langle \sigma^x\rangle(\omega)|^2$. The Fourier transform was performed for times up to $t=480$. The $E_8$ masses as well as gaps between them, denoted by $m_{ij} = m_i - m_j$, are marked in green. Unlabeled peaks correspond to multi-particle states that carry non-vanishing single-particle momenta.}
    \label{fig:power_spectrum}
\end{figure}

Figure~\ref{fig:power_spectrum} displays the power spectrum $|\langle \sigma^x\rangle(\omega)|^2$ generated from Fig.~2 in the main text. As the quench Hamiltonian lies in the $E_8$ scaling region, the excitations are described by the $E_8$ theory. Thus, the $E_8$ masses and gaps between them appear as peaks in the power spectrum of the longitudinal magnetization $|\langle \sigma^x \rangle (\omega) |^2$.

\end{document}